\documentclass[twocolumn]{aastex63}
\usepackage{comment}

\usepackage{amsmath}
\usepackage{mathrsfs} 
\usepackage{graphicx}
\usepackage{ulem}
\usepackage{soul}
\usepackage{xcolor}
\usepackage{macros_vdb}

\received{August 12, 2020}
\revised{September 11, 2020}
\accepted{September 15, 2020}

\shorttitle{Evolution of the Globular Cluster System in NGC 1052-DF2}
\shortauthors{Dutta Chowdhury et al.}

\begin{document}

\title{On the Evolution of the Globular Cluster System in NGC 1052-DF2: Dynamical Friction, Globular-Globular Interactions and Galactic Tides}
\correspondingauthor{Dhruba Dutta Chowdhury}
\email{dhruba.duttachowdhury@yale.edu}

\author[0000-0003-0250-3827]{Dhruba Dutta Chowdhury}
\affil{Department of Astronomy, Yale University \\
52 Hillhouse Avenue, New Haven, CT-06511, USA}

\author[0000-0003-3236-2068]{Frank C. van den Bosch}
\affil{Department of Astronomy, Yale University \\
52 Hillhouse Avenue, New Haven, CT-06511, USA}

\author[0000-0002-8282-9888]{Pieter van Dokkum}
\affil{Department of Astronomy, Yale University \\
52 Hillhouse Avenue, New Haven, CT-06511, USA}

\begin{abstract}
  The ultra-diffuse galaxy NGC 1052-DF2 has an overabundance of luminous globular clusters (GCs), and its kinematics is consistent with the presence of little to no dark matter. As the velocity dispersion among the GCs is comparable to the expected internal dispersions of the individual GCs, the galaxy might be highly conducive to GC-GC merging. If true, this could explain the puzzling luminosity function of its GCs. Here, we examine this possibility by re-simulating three of our earlier simulations of the GC system \citep[][]{duttachowdhury19}, where the GCs were modeled as single particles, with live GCs. Somewhat surprisingly, we infer a low merger rate of $\sim 0.03 \Gyr^{-1}$. The main reason is that the GCs are too dense for tidal shock capture, caused by impulsive encounters among them, to operate efficiently (we infer a tidal capture rate of only $\sim 0.002 \Gyr^{-1}$). Therefore, whatever mergers occur are driven by other mechanisms, which we find to be captures induced by dynamical friction and compressive tides from other GCs. The low merger rate inferred here makes it unlikely that the unusually large luminosities of the GCs can be explained as a result of past GC-GC mergers. Our simulations also indicate that, if NGC 1052-DF2 is indeed largely devoid of dark matter, its tidal field is too weak to induce any significant mass loss from the GCs. Therefore, in such a scenario, we predict that it is improbable for the GCs to reveal tidal features, something that can be tested with future deep observations. 
\end{abstract}


\section{Introduction}\label{intro}

The discovery of NGC 1052-DF2 \citep[hereafter DF2,][]{vandokkum18a} and NGC 1052-DF4 \citep[hereafter DF4,][]{vandokkum19} has revealed the existence of a puzzling population of globular cluster (GC) rich, dark matter deficient galaxies. Not only do these galaxies have an overabundance of luminous GCs \citep{vandokkum18b, vandokkum19}, but their kinematics are also consistent with the presence of little to no dark matter \citep{vandokkum18c, washerman18, vandokkum19}. The shift in the GC luminosity function to higher luminosities than usual is statistically significant and is not due to observational bias towards detecting more luminous GCs (see Shen et al. 2020, in prep). While the association of DF2 with the NGC 1052 group at $20 \Mpc$ and the robustness of its dynamical mass, inferred from GC kinematics, have been contested in several studies \citep{hayashi18, martin18, laporte19, trujillo19, nusser19a, lewis20}, both \citet{vandokkum18d} and \citet{blakeslee18} have independently confirmed its distance to be $19-20 \Mpc$. More importantly, using stellar kinematics, \citet{danieli19} and \citet{emsellem19} have validated its low dark-to-stellar mass ratio (at least within the optical extent).

In the standard paradigm of galaxy formation, a relatively massive dark matter halo is a prerequisite for cold gas to collapse and form stars. Therefore, how such large and diffuse galaxies (both DF2 and DF4 belong to the class of ultra-diffuse galaxies) with little to no dark matter content came into being is a puzzle. \citet{ogiya18} and \citet{nusser20} have proposed that these galaxies formed in more massive progenitor halos, which were then tidally heated and stripped in the NGC 1052 group environment, giving rise to dark matter depleted systems. However, such models do not address the origin of the overabundance of luminous GCs. In fact, since the distribution of GCs is typically more extended than the stellar body of the host galaxy, stripping is likely to result in a smaller rather than a larger specific frequency. An alternative scenario, due to \citet{silk19}, involves a high-velocity collision between two gas-rich galaxies that causes a spatial offset between their dark and baryonic components. The collision also triggers globular cluster formation, leading to a high specific frequency of GCs. That dark matter deficient galaxies can form in this way has also been shown by \citet{shin20}. 

Irrespective of how they form, the GC-rich, low-mass systems DF2 and DF4 present a unique environment for GC evolution. While the dynamics of the GCs can be used to constrain possible mass models \citep{nusser18, duttachowdhury19}, they are also interesting in their own right. In \citet[hereafter Paper~I]{duttachowdhury19}, we studied the dynamical evolution of the GC system in DF2 for a baryon-only mass model. Using $N$-body simulations, we showed that due to a cored stellar density profile, dynamical friction on the GCs is significantly reduced in the central region of the galaxy \citep[a phenomenon known as core-stalling, see also][]{hernandez98, read06, inoue09, inoue11, petts15, petts16, kaur18}. Paper~I also revealed frequent GC-GC scattering, which, together with core-stalling, prevents the GCs from sinking to the galaxy center. 

A shortcoming of Paper~I was that each GC was modeled as a single particle (i.e., a `hard sphere'). Consequently, we were unable to account for potential GC-GC mergers, which have been previously studied in the in the context of nuclear star cluster formation \citep[e.g.,][]{tremaine75, oh00, capuzzo-dolcetta08a, capuzzo-dolcetta08b, bekki10, hartmann11, arca-sedda14, gnedin14}, the evolution of disk GCs in the Milky Way \citep[e.g.,][]{khoperskov18,mastrobuono-battisti19} and that of stellar super-clusters \citep[e.g.,][]{kroupa98}, and the formation of ultra-compact dwarfs \citep[e.g.,][]{fellhauer02, bekki04}. Two gravitationally bound systems are likely to merge when their relative speed is lower than (or of the same order as) their internal dispersions \citep{binney08}. Since the velocity dispersion of the GC system in DF2 is comparable to the expected dispersions of the individual GCs (see Section~\ref{setup}), DF2's environment might be highly conducive for GC-GC mergers. It is tempting, therefore, to explain the extreme luminosities of the GCs in DF2, which are brighter than usual \citep[][]{vandokkum18b}, as being the outcome of such mergers. 

In this paper, we explore the dynamical evolution of the GCs in DF2 by modeling them as live $N$-body systems, rather than as hard spheres. As in Paper~I, we evolve the GC population in a live baryon-only model of the galaxy for a duration of 10 Gyr, starting from equilibrium initial conditions that match the observational constraints. In addition to focusing on GC-GC mergers, we examine the impact of the tidal field of DF2 and that of the other GCs on the mass and structural evolution of a GC. We also compare the orbital decay of the live GCs to that of the corresponding hard spheres. 

We emphasize that these simulations are needed to obtain a reliable estimate of the GC-GC merger rate in DF2. One may be inclined to estimate the merger rate from kinetic theory, equating the mean free path of a GC to $(n \sigma_{\rm tc})^{-1}$ with $n$ the number density of GCs and $\sigma_{\rm tc}$ the (velocity dependent) cross-section for tidal shock capture \citep[cf.][]{mamon92, makino97}. However, we caution that this does not account for dynamical friction, which plays an important role in that it causes the GCs to congregate near the code radius of DF2 due to core-stalling. Indeed, as we demonstrate in Section~\ref{discussion}, tidal shock capture is not the primary mechanism driving GC-GC mergers in DF2, and, therefore, such an analytic calculation grossly underpredicts the merger rate.

This paper is organized as follows. Section~\ref{setup} discusses how the simulations are set up using the observational constraints on the GCs in DF2. The results of our simulations are presented in Section~\ref{results}, followed by a detailed discussion of the inferred GC-GC merger rate in Section~\ref{discussion}. We summarize our findings in Section~\ref{summary}. 

\section{Simulation Setup} \label{setup}

\begin{table}
   \centering
   \begin{tabular}{| c | c | c | c | c | c |} 
   \hline
    Id & $M_{\rm GC}$ & $r_{\rm h,proj} (a)$ & $\sigma_{\rm 3D}$ & $\sigma_{\rm LOS}$ & $N$ \\
    & ($\Msun$) & ($\pc$) & ($\kms$) & ($\kms$) & \\
    \hline  
    39 & $7.3 \times 10^5$ & 7.5 & 12.8 & 7.3  & 36500 \\
    59 & $5.0 \times 10^5$ & 6.5 & 11.3 & 6.5  & 25000 \\
    71 & $5.5 \times 10^5$ & 6.7 & 11.7 & 6.7  & 27500 \\
    73 & $1.5 \times 10^6$ & 6.4 & 19.8 & 11.3 & 75000 \\
    77 & $9.6 \times 10^5$ & 9.4 & 13.1 & 7.4  & 48000 \\
    85 & $6.6 \times 10^5$ & 5.2 & 14.6 & 8.3  & 33000 \\
    91 & $6.6 \times 10^5$ & 8.4 & 11.5 & 6.5  & 33000 \\
    92 & $8.0 \times 10^5$ & 4.3 & 17.7 & 10.1 & 40000 \\
    98 & $4.2 \times 10^5$ & 5.4 & 11.4 & 6.5  & 21000 \\
    101 & $3.8 \times 10^5$& 4.8 & 11.5 & 6.6  & 19000 \\
    \hline
    \end{tabular}
    \caption{Columns 1 through 3 list the IDs, masses and projected half-light radii (scale radii of the corresponding Plummer spheres) of the 10 spectroscopically confirmed GCs in DF2 considered in this paper. For each GC, columns 4, 5 and 6 list the 3D velocity dispersion inside the 3D half-mass radius, the LOS velocity dispersion inside the projected half-mass radius, and the total number of star particles used in its $N$-body representation, respectively. See text for details.}
    \label{gc_prop} 
\end{table}

We model the diffuse stellar component of DF2 as a spherically symmetric, isotropic system in equilibrium. Assuming a total mass of $M=2 \times 10^{8} \Msun$ and a distance of $20 \Mpc$ \citep{vandokkum18a}, the three-dimensional (3D) density profile of the stars is inferred from the observed \cite{sersic68} surface brightness profile (with S\'ersic index, $n=0.6$ and effective radius, $R_\rme=2.2 \kpc$) using the inverse Abel transformation. After that, Eddington inversion is used to obtain the corresponding ergodic distribution function (DF), $f(\varepsilon)$, where $\varepsilon$ is the negative of the energy per unit mass of a star particle (for more details, see Paper~I). The DF, thus obtained, is then used to draw positions and velocities for $10^{7}$ star particles, each having a mass of $20 \Msun$. This mass resolution is a factor of 10 better than that adopted in Paper~I.

Each of the ten spectroscopically confirmed GCs in DF2 is set up as a spherically symmetric, isotropic \citet{plummer11} sphere in equilibrium, whose DF is given by 
\begin{equation}
f(\varepsilon) = F \, \varepsilon^{7/2} \ . 
\label{plummer_df}
\end{equation}
The constant, $F$, depends on the mass, $M_{\rm GC}$, and scale radius, $a$, of the Plummer sphere and is derived from the constraint that $\int \int f(\varepsilon) \, \rmd^3{\bf v}\, \rmd^3 {\bf r} = M_{\rm GC}$. The resulting density profile is given by
\begin{equation}
\rho(r)=\frac{3M_{\rm GC}}{4 \pi a^3} \left( 1+r^2/a^2 \right)^{-5/2} \ .
\end{equation}
The initial mass of each GC is set equal to that inferred from its observed luminosity \citep{vandokkum18b}, using a constant mass-to-light ratio of $1.8$. The initial scale radius of each GC is set equal to its observed projected half-light radius\footnote{Assuming a constant mass-to-light ratio, the half-light radius is equal to the half-mass radius, and for a Plummer sphere, the projected half-mass radius is equal to the scale radius, $a$.} \citep{vandokkum18b}. Initial phase-space coordinates of star particles, each of mass $20 \Msun$, for each GC, are sampled from the DF of Equation~\ref{plummer_df}. Table~\ref{gc_prop} lists the properties of all 10 GCs, including their masses, projected half-light radii (scale radii), 3D velocity dispersions inside the respective 3D half-mass radii, $\sigma_{\rm 3D}$, line-of-sight (LOS) velocity dispersions inside the respective projected half-mass radii, $\sigma_{\rm LOS}$, and the number of particles, $N$, used to represent each GC. Note that $\sigma_{\rm 3D}$ and $\sigma_{\rm LOS}$ are inferred from the DF of Equation~\ref{plummer_df}, and since all the GCs are self-similar, we have $\sigma_{\rm LOS} \simeq 0.57 \sigma_{\rm 3D}$ in each case.

\begin{figure}
    \centering
    \includegraphics[width=0.45\textwidth]{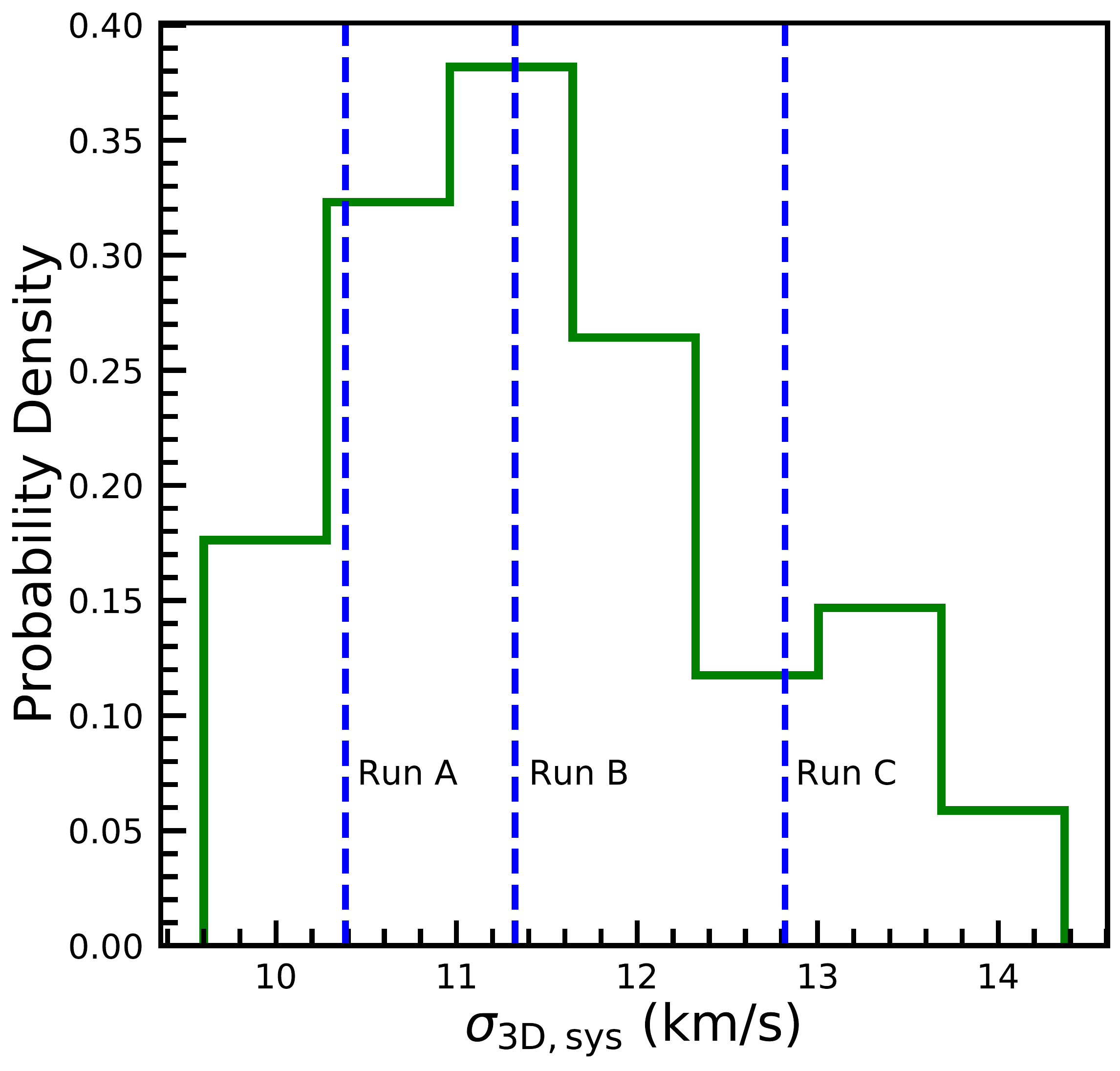}
    \caption{Probability distribution for the 3D velocity dispersion of the GC system, $\sigma_{\rm 3D, sys}$, at $t=0$, as obtained from the 50 multi-GC realizations described in Paper~I (green histogram). The blue, dashed, vertical lines denote the realizations that are chosen for re-simulation with live GCs and correspond to the $16^{\rm th}$, $50^{\rm th}$, and $84^{\rm th}$ percentiles of the $\sigma_{\rm 3D, sys}$ distribution. Due to computational limitations, it is not feasible to re-simulate all 50 realizations.}
    \label{fig:gc_system_dispersion}
\end{figure}

\begin{figure*}
    \centering
    \includegraphics[width=0.9\textwidth]{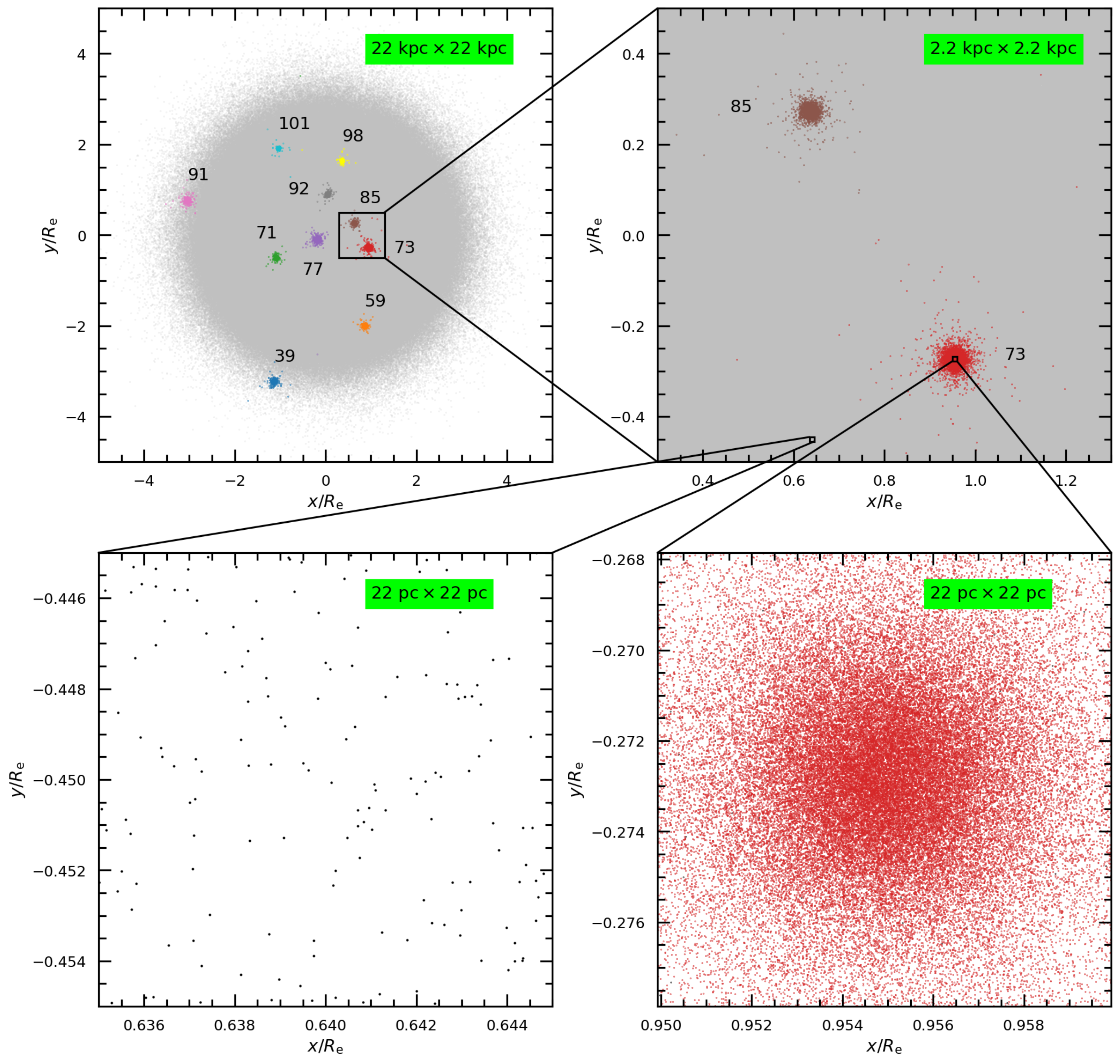}
    \caption{The upper-left panel shows the $N$-body representation of DF2 and its GCs projected on the sky-plane at $t=0$, covering an area of $22 \kpc \times 22 \kpc$. The star particles that belong to the galaxy are shown in gray, and those associated with the GCs are displayed with different colors, as indicated. By having live GCs (with $19000-75000$ particles, depending on the GC mass) within a live galaxy (with $10^{7}$ particles), our simulations resolve a vast range of densities and scales. This is illustrated by zooming into a region of area $2.2 \kpc \times 2.2 \kpc$ that contains GCs 73 and 85 (upper-right panel), and further zooming into two regions of area $22 \pc \times 22 \pc$, one that does not contain any GC (lower-left panel) and another centered on GC 73 (lower-right panel). Note the huge density contrast between these two zoom-ins.}
    \label{fig:snapshot}
\end{figure*}

\begin{figure*}
    \centering
    \includegraphics[width=0.9\textwidth]{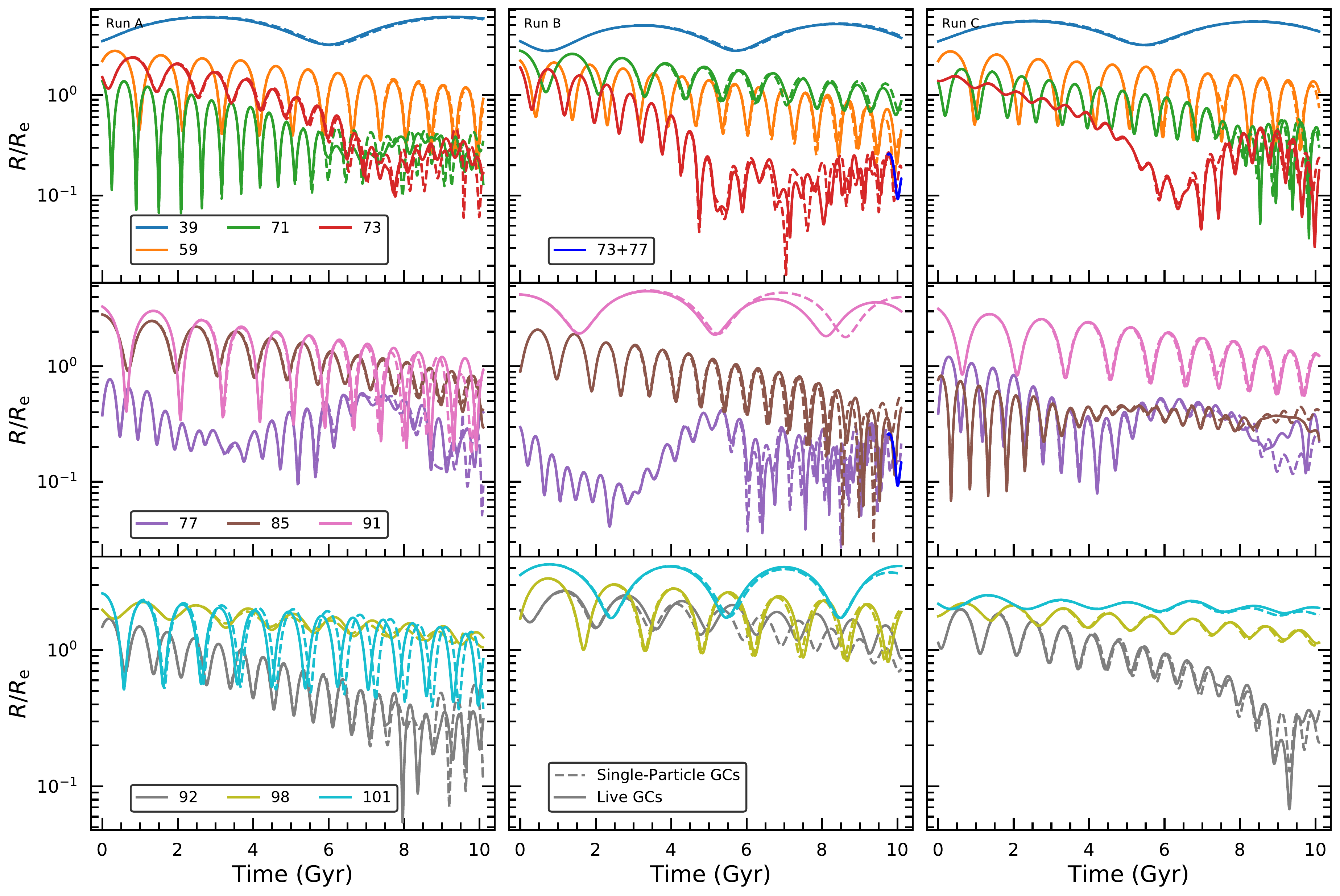}
    \caption{Orbital Evolution of Live and Single-Particle GCs. Solid lines indicate the evolution of the galactocentric distance, $R$, of each live GC, in units of the effective radius of DF2, $R_\rme$, in Runs A (left-hand column), B (central column), and C (right-hand column). The orbital evolution of the GCs in Runs A', B', and C', which have the same initial conditions as Runs A, B, and C, respectively, but where the GCs are represented as hard spheres, is shown with dashed lines. In each column, the GCs are divided into three subsets (top, middle, and bottom panels) for clarity. Overall, the evolution of live and single-particle GCs are in very good agreement. The small differences that occur in the later stages of the evolution are due to the transfer of relative orbital energy of a GC pair to internal energy of the live GCs. In Run B, GCs 73 and 77 merge together at $t=9.75 \Gyr$, and the orbit of the merged remnant is indicated by the blue curves in the top and middle panels of the central column. See text for more details.}
    \label{fig:orb_comp}
\end{figure*}

In Paper~I, we used the observed projected positions and LOS velocities of the GCs \citep{vandokkum18a,vandokkum18c}, both measured with respect to the galaxy center, as constraints to make 50 realizations for the GC system. This was done by first determining the 3D number density profile of the GCs from their projected number density (fitted with a S\'ersic profile of index, $n=1$ and 2D half-number radius, $R_{\rm half, GC}=1.3\ R_\rme$) using the inverse Abel transformation. Next, Eddington inversion was used to calculate the corresponding ergodic DF of the GC system by assuming it to be in equilibrium with the stellar potential. The DF, thus obtained, was then used to sample GC positions along the LOS and velocity components perpendicular to the LOS. For more details, see Paper~I.

Figure~\ref{fig:gc_system_dispersion} shows the probability distribution for the 3D velocity dispersion of the GC system, $\sigma_{\rm 3D, sys}$, in the 50 realizations presented in Paper~I. Note that the typical velocity dispersion of the GC system ($\sim 10-14 \kms$) is comparable to, and in some cases, even lower than the internal velocity dispersions of the individual GCs, listed in Table~\ref{gc_prop} ($\sim 11 - 20 \kms$). As noted in Section~\ref{intro}, such a situation is conducive for GC-GC mergers. 

\begin{figure*}
    \centering
    \includegraphics[width=0.9\textwidth]{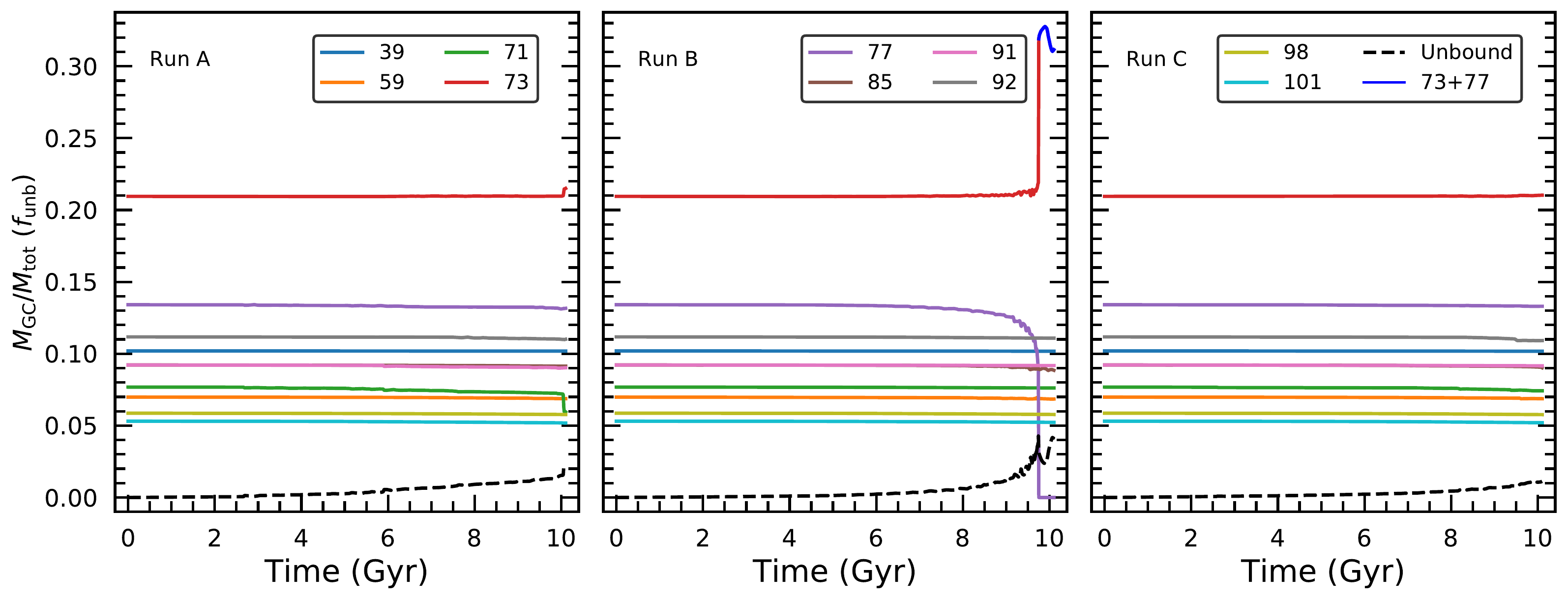}
    \caption{Evolution of the bound mass of each GC, $M_{\rm GC}$, normalized by the total mass of all the GCs at $t=0$ ($M_{\rm tot}$), in Runs A (left-hand panel), B (middle panel), and C (right-hand panel). Each GC is represented with a different color, as indicated. The fraction, $f_{\rm unb}$, of $M_{\rm tot}$ that is no longer bound to any of the GCs is indicated by the black, dashed curves. Overall, the GCs are very stable to tidal perturbations from the galaxy and other GCs. They undergo little mass evolution over 10 Gyr, except during GC-GC mergers when two GCs are strongly affected by each other's tidal field and experience significantly more mass loss/gain (GCs 71 and 73 in Run A; GCs 77 and 73 in Run B). After the merger between GCs 73 and 77 occurs in Run B, the red curve in the middle panel is continued with a blue curve, reflecting the mass evolution of the merged remnant.}
    \label{fig:bound_fractions}
\end{figure*}

Since $\sigma_{\rm 3D, sys}$ indicates how fast the GCs are moving, on average, with respect to the galaxy center, it can be expected to be indicative of how many GCs are likely to merge in a particular realization. Therefore, in order to roughly sample, in some quantitative measure, the expected frequency of GC-GC mergers in DF2, we re-simulate, with live GCs, the realizations corresponding to the $16^{\rm th}$, $50^{\rm th}$, and $84^{\rm th}$ percentiles of the $\sigma_{\rm 3D, sys}$ distribution (indicated by the blue, dashed, vertical lines in Figure~\ref{fig:gc_system_dispersion}). In what follows, we refer to these simulations as Runs A, B, and C, respectively. To compare the orbital evolution of the live GCs with that of hard spheres, we also re-simulate the same realizations with single-particle GCs (hereafter Runs A', B', and C')\footnote{These simulations yield results that are indistinguishable from the corresponding simulations presented in Paper~I even though here, we use an order of magnitude more particles to represent the galaxy.}.  

After initializing the positions and velocities of the star particles that make up the galaxy and the individual GCs in isolation, we place the GCs at their respective positions within the galaxy and add the corresponding orbital velocity vectors to that of the individual GC particles. The upper-left panel of Figure~\ref{fig:snapshot} shows the $N$-body representation of DF2 and its GCs projected on the sky-plane at $t=0$, which is the same for all three live-GC simulations. The zoom-ins illustrate the huge dynamic range in scales and densities covered by our simulations.

All simulations are run forward in time for $10 \Gyr$ with the code GADGET-2 \citep{springel05}. In GADGET-2, the gravitational force between two particles is softened with a spline, such that the force is exactly Newtonian beyond $2.8 \epsilon$, where $\epsilon$ is the equivalent Plummer softening. We also simulate the most and least massive GCs and the stellar body of DF2 in isolation with different values of $\epsilon$ ($0.1-1 \pc$). For $\epsilon = 0.4\pc$, we find that both the stellar body of DF2 and the GCs in isolation remain in stable equilibrium for at least their respective half-mass relaxation times \citep[computed using][]{spitzer69}. We, therefore, adopt this value of the softening length in Runs A, B, and C for all particles, independent of whether they belong to DF2 or one of the GCs. In Runs A', B', and C', we adopt $\epsilon=10 \pc$ for all particles (stars and GCs). In Paper~I, this was found to be the optimal softening length for the single-particle GCs and perfectly adequate to model the stellar body of DF2 as well. A Barnes-Hut oct-tree \citep{barnes86} with an opening angle of $0.7$ is used for gravitational force calculations and the time step, $\Delta t$, taken by a particle is determined using the criterion, $\Delta t=\sqrt{ 2 \eta \epsilon/|{\bf a}|}$. Here, {\bf a} is the instantaneous acceleration of the particle, and $\eta$ controls the accuracy of time integration. We set $\eta=0.002$ in all runs.

\section{Results}\label{results}

For the live-GC simulations, determining the membership of the GCs (i.e., which particle is bound to which GC) is a non-trivial exercise. Once the membership of the GCs in a particular snapshot has been ascertained, the member particles of a GC can be used to determine its center of mass position and velocity, bound mass, and structural properties. Therefore, before presenting the simulation results, we briefly outline the steps taken to accomplish this task.
\begin{enumerate}
    \item For each GC, in each snapshot, we initialize its membership with the particles that belonged to it at $t=0$ and find the position and velocity of the center of mass of this collection.
    
    \item For each GC, this collection of particles is then fed to a tree code, with the same softening length and opening angle as that used in the live-GC simulations, along with those particles that do not belong to this collection (but belonged to any one of the ten GCs at $t=0$) but with their masses set to zero. 
    
    \item The potential calculated by the tree code for each particle is used to obtain its binding energy in the center of mass frame of each GC, as determined in the previous iteration (or in step 1 for the zeroth iteration). The membership of the GCs is updated by assigning each particle to the GC with respect to which it has the most negative binding energy. The particles that have positive binding energy with respect to every GC are not assigned to any and constitute the collection of unbound particles. 
    
    \item For each GC, from its collection of bound particles determined in step 3, the $50 \%$ most bound ones are used to update the position and velocity of its center of mass. 
\end{enumerate}
Steps 2, 3, and 4 are repeated with each GC's collection of bound particles, as determined in the previous iteration, until the relative separation between the positions and velocities of the center of masses obtained in two successive iterations converges to better than $1 \times 10^{-3}$ for every GC.  

\subsection{Orbital Evolution of Live and Single-Particle GCs} \label{orbits}

Figure~\ref{fig:orb_comp} shows the orbital evolution of the GCs in DF2. From left to right, the three different columns depict the galactocentric distances of the ten GCs as a function of time for the three different initial condition setups described in Section~\ref{setup}. In each column, the top, middle, and bottom panels show the results for different GC subsets (to avoid overcrowding), with each GC represented by a different color, as indicated. For each GC, we depict its orbital evolution in both the simulations where it is live (solid lines; Runs A, B, and C) and where it is represented as a hard sphere (dashed lines; Runs A', B', and C').

Overall, the orbital evolution of the live and single-particle GCs show very good agreement. Small differences occur in the later stages of the evolution ($t \gtrsim 5 \Gyr$) due to the presence of extra degrees of freedom in the live GC simulations, pertaining to the internal motion of the GCs. As the GCs sink in, due to dynamical friction, and come closer together at later times, GC-GC interactions become important. Together with reduced dynamical friction in the galactic core, these interactions keep the GCs afloat, preventing them from sinking to the center of the galaxy (see Paper~I). In the case of live GCs, GC-GC interactions also transfer orbital energy from the relative motion of a GC pair to internal energy of the GCs, causing the GC orbits to deviate from that in the corresponding simulations with single-particle GCs. If this transfer of energy is sufficiently large or continues for a sufficiently long time, it leads to a GC-GC merger. 

Somewhat surprisingly, even though DF2 was purported to be conducive to GC-GC merging, we only find a single, complete merger event in our simulations. This merger, which occurs in Run B and involves GCs 73 and 77, happens towards the very end of the simulation, at $t=9.75 \Gyr$. The blue curves in the top and middle panels of the central column in Figure \ref{fig:orb_comp} show the orbital evolution of the merged remnant, and it is joined to the red (GC 73) and magenta curves (GC 77) at $t=9.75 \Gyr$, the time when it is no longer possible to identify the two GCs as separately bound systems. Note that the merged remnant continues to orbit near the core radius of DF2 (roughly $0.2-0.3\ R_\rme$) and does not sink to the galaxy center. Section~\ref{merger} discusses this merger event in more details.

\subsection{Mass Evolution of Live GCs} \label{mass}

\begin{figure}
    \centering
    \includegraphics[width=0.45\textwidth]{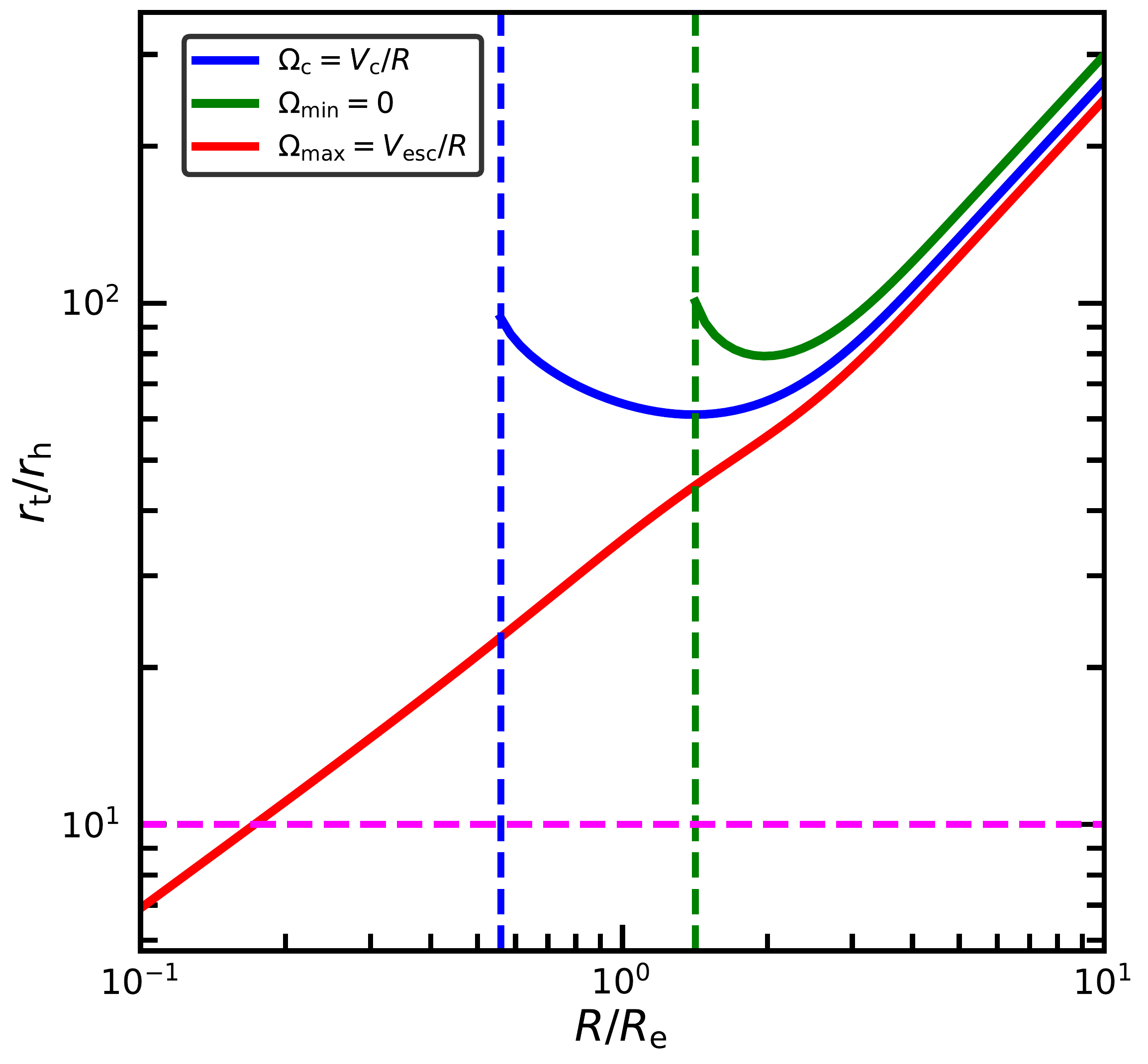}
    \caption{The tidal radius, $r_\rmt$, of a typical GC in DF2, in units of its 3D half-mass radius, $r_\rmh$, as a function of its galactocentric distance, $R$, in units of the effective radius of DF2, $R_\rme$.  The solid, blue curve shows the tidal radius for a circular orbit ($\Omega=\Omega_\rmc=V_\rmc/R$), while solid, red and green curves depict the instantaneous tidal radius for maximum ($\Omega_{\rm max}=V_{\rm esc}/R$) and minimum ($\Omega_{\rm min}=0$) possible angular velocities at a particular $R$, respectively. The dashed, vertical, blue (green) line indicates the galactocentric distance inside which the tidal radius is infinite for $\Omega=\Omega_\rmc\ (\Omega_{\rm min})$. The dashed, magenta, horizontal line highlights the GC radius that encloses $99 \%$ of its total mass. Except for rare pericentric passages close to the center ($R<0.2\ R_\rme$) on highly eccentric orbits ($\Omega$ close to $\Omega_{\rm max}$), mass loss due to galactic tides is insignificant.}
    \label{fig:gc_tidal_radius}
\end{figure}

\begin{figure*}
    \centering
    \includegraphics[width=0.9\textwidth]{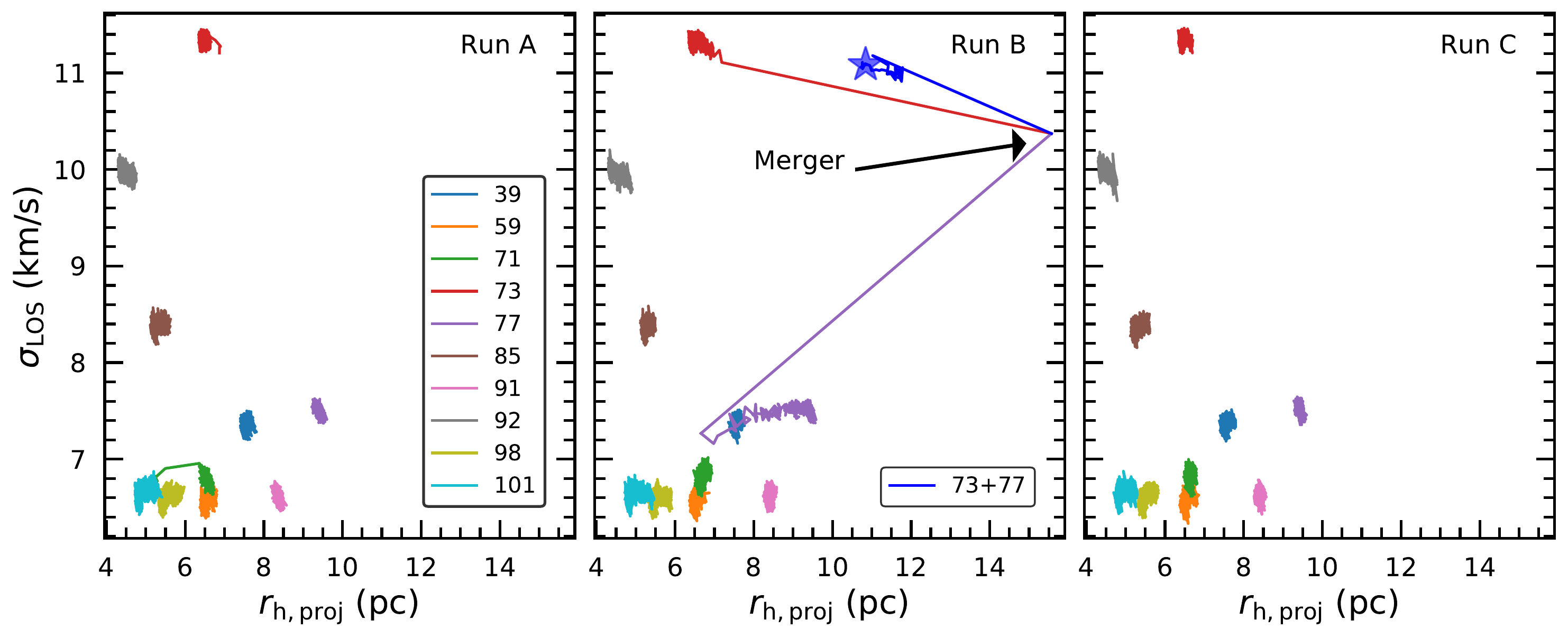}
    \caption{Evolution of the GCs in the $r_{\rm h,proj}-\sigma_{\rm LOS}$ plane, where $r_{\rm h,proj}$ is the projected half-mass radius of a GC and $\sigma_{\rm LOS}$ is its LOS velocity dispersion inside $r_{\rm h,proj}$. Results are shown for Runs A (left-hand panel), B (middle panel), and C (right-hand panel). Different colors represent different GCs, as indicated. Overall, there is remarkably little evolution in the structural parameters of the GCs. A clear exception is the merger between GCs 73 and 77 in Run B (red and magenta curves in the middle panel), which produces the remnant whose evolution is indicated in blue, with the asterisk marking its final state at the end of $10 \Gyr$.}
    \label{fig:proj_rad_los_dis}
\end{figure*}

Figure~\ref{fig:bound_fractions} shows the evolution of the bound mass of each live GC, $M_{\rm GC}$, normalized by the total mass of all the GCs at $t=0$, $M_{\rm tot} = \sum_{i=1}^{10} M^{i}_{\rm GC}$. Different panels correspond to different runs, as indicated. Note that at any given time, the bound mass of a GC consists of star particles that belonged to it initially and are still bound (self mass) as well as star particles that belonged to the other GCs at $t=0$ but are now bound to this particular GC (accreted mass). The black, dashed curves in each panel indicate the fraction, $f_{\rm unb}$, of $M_{\rm tot}$ that is no longer bound to any of the GCs. 

In Runs A, B, and C, after 10 Gyr of evolution, $f_{\rm unb}$ is $2 \%$, $4 \%$, and $1 \%$, respectively. Such low values of $f_{\rm unb}$ indicate that the GCs are very stable to mass loss induced by tidal perturbations from the galaxy. For a single GC on a circular orbit in DF2, its tidal radius, $r_\rmt$, is the distance to the Lagrange point, $L_3$, from the center of the GC, and is given by the root of the equation
\begin{equation}
\frac{GM(R-r_\rmt)}{(R-r_\rmt)^2}-\frac{GM(R)}{R^2}-\frac{GM_{\rm GC}(r_\rmt)}{r_\rmt^2}+\Omega_\rmc^{2}r_\rmt=0 \ .
\label{t_radius}
\end{equation}
Here, $R$ is the galactocentric distance of the GC, $M(R)$ is the mass of DF2 enclosed within $R$, $M_{\rm GC}(r)$ is the GC mass enclosed within the cluster-centric radius, $r$, and $\Omega_\rmc$ is its angular velocity. The density profile of the galaxy is given by the deprojected S\'ersic profile, as discussed in Section~\ref{setup}, and the GC has a Plummer density profile with mass, $M_{\rm GC}=7.2 \times 10^{5} \Msun$ and scale radius, $a=6.5 \pc$, which are the initial averages of the respective quantities for the ten GCs in DF2.

The solid, blue curve in Figure~\ref{fig:gc_tidal_radius} shows $r_\rmt$ as a function of $R$, obtained by solving Equation~\ref{t_radius}. The dashed, blue, vertical line indicates the galactocentric distance inside which $L_3$ ceases to exist for a circular orbit, and the tidal radius is infinite. This is due to the cored density profile of DF2, which causes tidal forces to become fully compressive at small $R$. For eccentric orbits, one can define the instantaneous tidal radius by replacing $\Omega_\rmc$ in Equation~\ref{t_radius} with the instantaneous angular velocity, $\Omega$, of the GC. At a given $R$, $\Omega<\Omega_\rmc\ (>\Omega_\rmc)$ indicates apocentric (pericentric) passages. The solid, red and green curves depict the instantaneous tidal radius as a function of $R$ for maximum ($\Omega_{\rm max}=V_{\rm esc}/R$) and minimum ($\Omega_{\rm min}=0$) possible angular velocities at that $R$, respectively. For $\Omega_{\rm min}$, $L_3$ ceases to exist (and the tidal radius is infinite) inside the galactocentric distance indicated by the dashed, green, vertical line. For $\Omega_{\rm max}$, the centrifugal force is maximum and effectively counters the compressive tidal force in the core. As a result, the tidal radius is always finite and continues to decrease with decreasing $R$. The dashed, magenta, horizontal line indicates $r_{99}$, the GC radius that encloses $99 \%$ of the total GC mass. Except for pericentric passages close to the center ($r<0.2\ R_\rme$) on highly eccentric orbits ($\Omega$ close to $\Omega_{\rm max}$), $r_\rmt$ is always much larger than $r_{\rm 99}$. Therefore, for the GCs in DF2, mass loss due to galactic tides is almost always insignificant (at least if DF2 is devoid of dark matter, as assumed here).

The tidal field of one GC on another is also not strong enough to cause significant mass evolution unless a pair of GCs is about to undergo a merger. For example, in Run B, maximum mass loss is experienced by GC 77. By the time of its merger to GC 73, $87 \%$ of its self mass is accreted onto GC 73, and the remaining $13 \%$ is not bound to any GC. During the same time, GC 73 loses only about $4 \%$ of its self mass, almost all of which is no longer bound to any other GC. However, the mass accreted from GC 77 more than compensates for this loss and increases its bound mass by about $52 \%$ compared to that at $t=0$ (red curve in the middle panel of Figure~\ref{fig:bound_fractions}). Post-merger, this red curve is continued with a blue curve, indicating the mass evolution of the merged remnant. Similarly, in Run A, towards the very end of the simulation, GCs 71 and 73 (green and red curves in the left-hand panel of Figure~\ref{fig:bound_fractions}, respectively) start getting strongly affected by each other's tidal field. GC 71, being the less massive of the two, loses mass to GC 73. As the merger has just begun, the mass exchange is less pronounced than that between GCs 73 and 77 in Run B. 

\begin{figure*}
    \centering
    \includegraphics[width=0.9\textwidth]{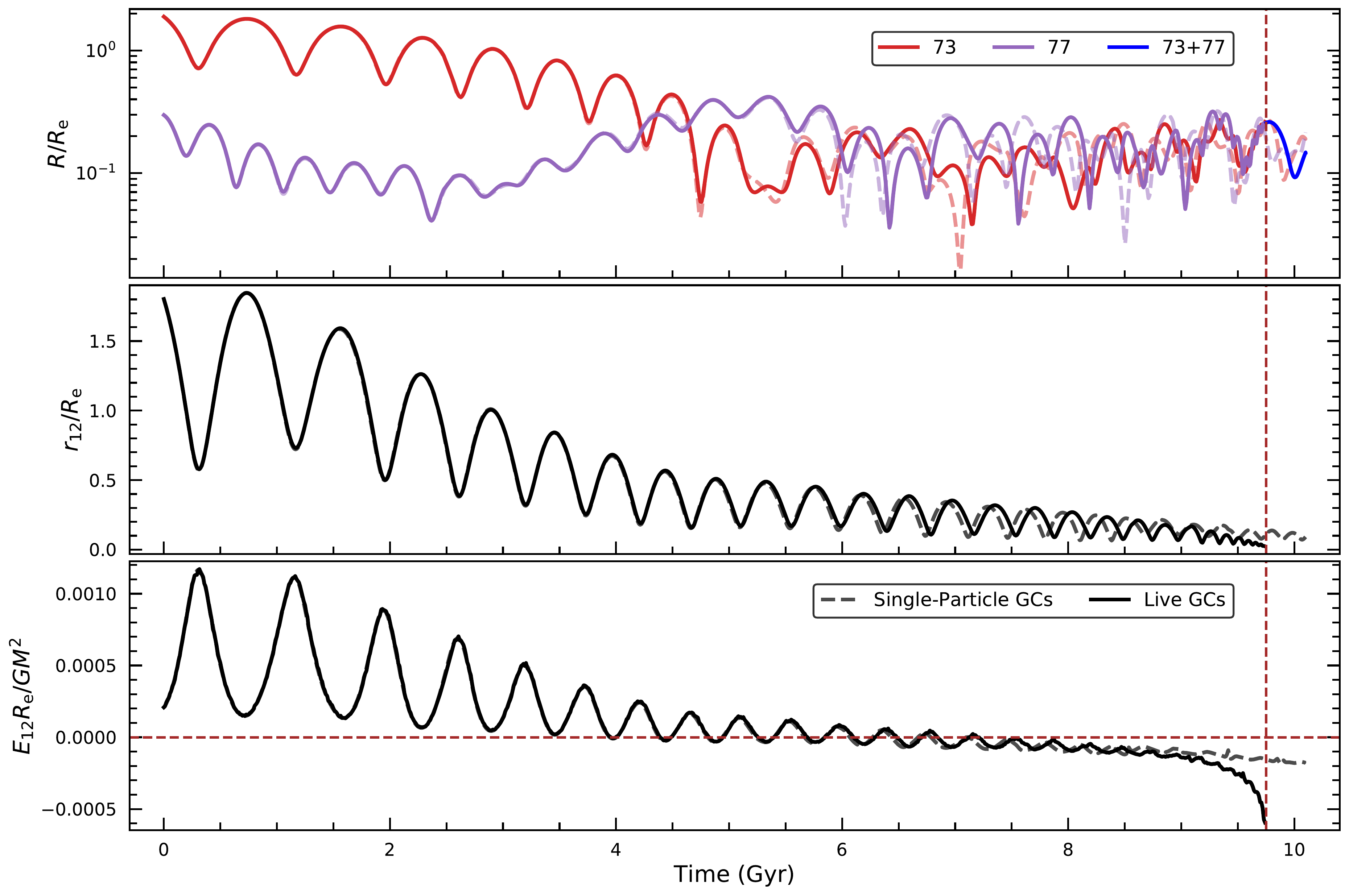}
    \caption{Evolution of GCs 73 and 77 in Runs B (solid curves) and B' (dashed curves). Red and magenta curves in the top panel show the evolution of the galactocentric distances of GCs 73 and 77, respectively, in units of the effective radius of DF2, $R_\rme$. The black curves in the middle and lower panels depict the evolution of the relative separation between the two GCs, $r_{\rm 12}$, in units of $R_\rme$, and their relative orbital energy, $E_{12}$, in units of $GM^{2}/R_e$, respectively. Here, $M$ is the mass of DF2 and $G$ is the universal gravitational constant. Loss of $E_{12}$ due to dynamical friction brings the GCs close together, allowing them to become bound. In the case of live GCs, after becoming bound, $E_{12}$ starts getting converted to internal energy of the GCs, eventually resulting in a merger at around $9.75 \Gyr$. In the upper panel, the galactocentric distance of the merged remnant, in units of $R_\rme$, is shown in blue and the time of the merger is denoted by the brown, dashed, vertical line. The brown, dashed, horizontal line in the lower panel indicates $E_{12}=0$.}
    \label{fig:merger_quan}
\end{figure*}

\begin{figure*}
    \centering
    \includegraphics[width=0.9\textwidth]{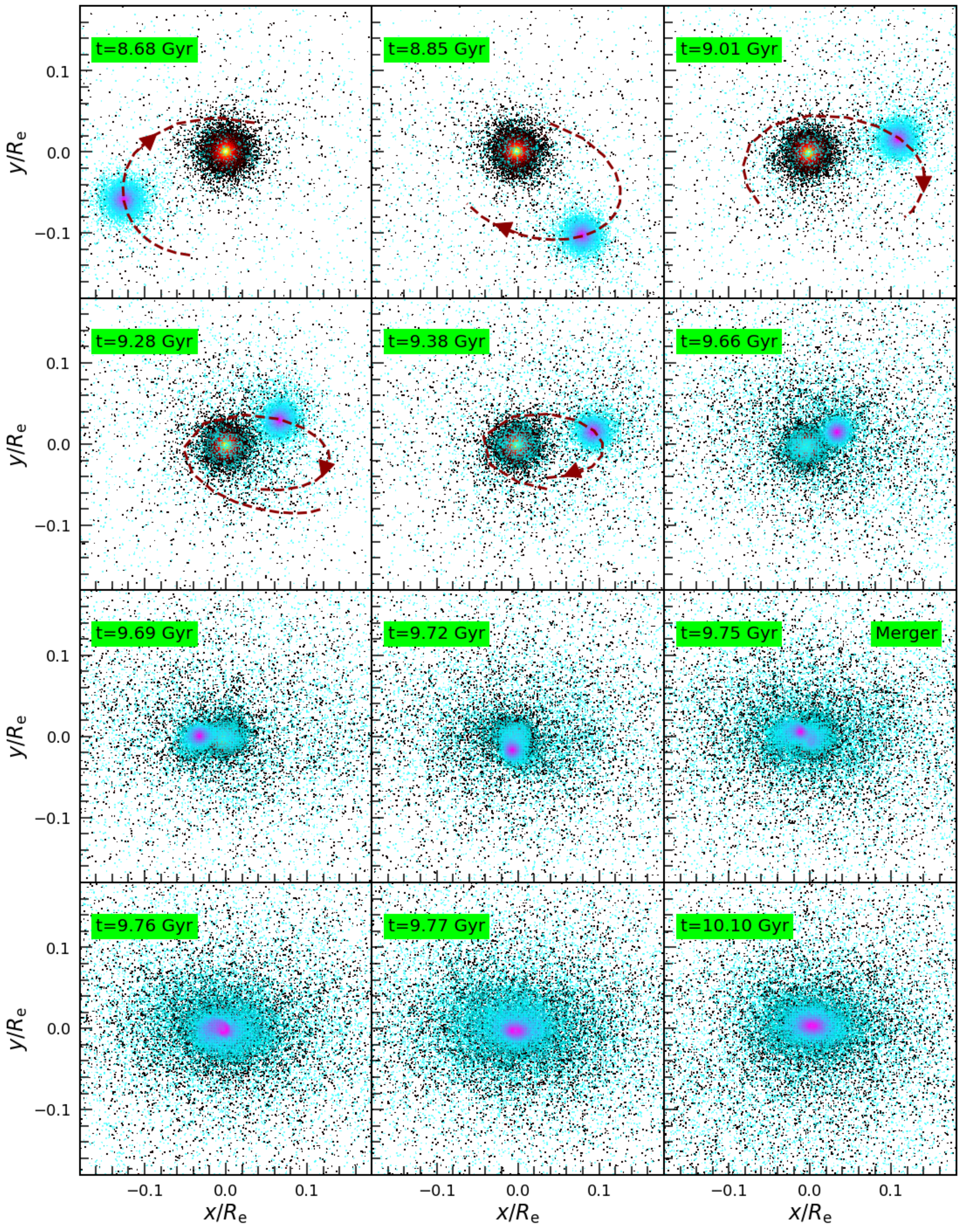}    
    \caption{Merger between GCs 73 and 77 in Run B. The projected density in the sky-plane of star particles that initially belonged to GCs 73 and 77, measured within pixels of area $2 \times 2\ \rm pc^2$, is depicted with the hot (varying from yellow to black with decreasing density) and cool (varying to magenta to cyan with decreasing density) colormaps, respectively, for a few snapshots before and after the merger. When the two GCs are sufficiently far apart, a portion of their relative orbit is also indicated with a brown, dashed curve. Over time, GC 77 gets closer to GC 73 and loses mass, the majority of which is accreted onto GC 73, resulting in a complete merger by $9.75 \Gyr$. Note the elongated nature of the final merged remnant.}
    \label{fig:merger_pic}
\end{figure*}

\subsection{Structural Evolution of Live GCs} \label{structure}

Figure~\ref{fig:proj_rad_los_dis} shows the evolution of the live GCs in the $r_{\rm h, proj}-\sigma_{\rm LOS}$ plane, where $r_{\rm h, proj}$ is the projected half-mass radius of a GC and $\sigma_{\rm LOS}$ is its LOS velocity dispersion inside $r_{\rm h, proj}$. Different panels correspond to different runs, and each GC is represented with a different color, as indicated. Overall, over 10 Gyr, there is remarkably little evolution in the structural parameters of the GCs, such that the curves for the individual GCs morph into little smudges. This indicates that, in general, the tidal field of the host galaxy and that of the other GCs have negligible effects on the structural evolution of a GC. 

The clear exception is the merger between GCs 73 and 77 in Run B. In this case, the two GCs are strongly affected by each other's tidal field. Before merging with GC 73, GC 77's $r_{\rm h,proj}$ and $\sigma_{\rm LOS}$ decrease by about $30 \%$ and $4 \%$, respectively (magenta curve in the middle panel). During the same time, GC 73's $r_{\rm h,proj}$ increases by about $12 \%$, and its $\sigma_{\rm LOS}$ decreases by about $2 \%$ (red curve in the middle panel). Post-merger, the red curve in the middle panel is continued with a blue curve, indicating the evolution of the merged remnant in the $r_{\rm h,proj}-\sigma_{\rm LOS}$ plane, and the blue star highlights its location at the end of $10 \Gyr$. In Run A, the merger between GCs 71 and 73 has just begun, so their evolution in the $r_{\rm h,proj}-\sigma_{\rm LOS}$ plane (green and red curves in the left-hand panel, respectively) is not as pronounced as that of GCs 73 and 77 in Run B.  

\subsection{A close-up look at the GC-GC Merger} \label{merger}

As mentioned above, across our three simulations only a single GC-GC merger occurs. Here, we describe this merger in some detail. The upper panel of Figure~\ref{fig:merger_quan} shows the galactocentric evolution of GCs 73 (red curve) and 77 (magenta curve) in Runs B and B'. The middle and lower panels show the evolution of the relative separation between the two GCs, $r_{12}$, and their relative orbital energy, $E_{12} = 0.5 \mu v_{12}^2 + W_{12}$, respectively. Here, $v_{12}$ is the relative velocity between the two GCs, $\mu$ is the reduced bound mass, and $W_{12}$ is the mutual gravitational potential energy. For two mass distributions with densities $\rho_{1}$ and $\rho_{2}$,
\begin{equation}
W_{12}=\frac{1}{2} \int \rho_{1}({\bf r})\Phi_{2}({\bf r}) d^{3} {\bf r} + \frac{1}{2} \int \rho_{2}({\bf r})\Phi_{1}({\bf r}) d^{3} {\bf r} \ ,
\label{pot_def}
\end{equation}
where $\Phi_{1}$ and $\Phi_{2}$ are the potentials due to 1 and 2, respectively. In the case of live GCs, $W_{12}$ is calculated by modeling GCs 73 and 77 as Plummer spheres, each with mass and scale radius equal to the instantaneous bound mass and projected half-mass radius of the GCs, respectively. In the case of single-particle GCs, Plummer spheres of scale radii equal to the softening length of Run B' ($10 \pc$) are used. In all panels, solid and dashed curves correspond to Runs B and B', respectively. 

GC 73, being the most massive globular cluster, experiences significant orbital decay due to dynamical friction and sinks to the galactic core in about $5 \Gyr$. GC 77's orbit also decays initially for about $2 \Gyr$, but as the orbit of GC 73 shrinks, and it gets closer to GC 77, GC 73 pulls GC 77 towards it and away from the galactic center for the next $3 \Gyr$ or so. The resulting decrease in the gravitational potential energy of the pair, which would otherwise appear as an increase in the relative kinetic energy of the GCs and cause them to drift apart again, is drained away to the galactic stars via dynamical friction. This can be inferred from the rapid, overall decreasing trend in $E_{12}$ for the first $5 \Gyr$, which leads to a similarly rapid, overall decrease in $r_{12}$. Note that during this period, the galactocentric evolution of both GCs and the evolution of $r_{12}$ and $E_{12}$ in Runs B and B' are in very good agreement. This is because the GCs are still sufficiently far apart, so their mutual attraction is not strong enough to distort their internal structure, i.e., transfer relative orbital energy to internal energy of the live GCs. Thus, they behave like single-particle systems for all practical purposes.

For the next $5 \Gyr$, both GCs 73 and 77 remain near the galactic core and continue to interact with each other. In both Runs B and B', dynamical friction keeps draining $E_{12}$, albeit at reduced efficiency (see Paper~I), allowing the GCs to become bound ($E_{12}<0$). After becoming bound, the GCs continue to get closer and closer and begin to interact more strongly. Consequently, in Run B, $E_{12}$ starts getting converted to internal energy of the GCs. Over time, this causes the evolution of $r_{12}$ and $E_{12}$ and the galactocentric evolution of both GCs to deviate more and more from that in Run B'. Eventually, at around $9.75 \Gyr$, the two GCs merge. In the upper panel of Figure~\ref{fig:merger_quan}, the evolution of the galactocentric distance of the merged remnant is shown in blue, and the time of merger is denoted by the brown, dashed, vertical line. Note that even after becoming bound, it takes an additional $2.5-3 \Gyr$ for the GCs to merge.

Figure~\ref{fig:merger_pic} shows several snapshots of the merger between GCs 73 and 77. The particle distributions are projected on the sky-plane and centered on the center of mass of all particles that initially belonged to GC 73. The projected density of all particles that initially belonged to GC 73, measured within pixels of area $2 \times 2\ \rm pc^2$, is depicted with the hot colormap (color changes from yellow to black with decreasing density). Similarly, the projected density of all particles that initially belonged to GC 77 is shown with the cool colormap (color changes from magenta to cyan with decreasing density). When the two GCs are sufficiently far apart, a portion of their relative orbit is also indicated with a brown, dashed curve. As the GCs get closer and begin to interact strongly with each other, relative orbital energy of the GC pair is converted to internal energy of the GCs (see also Figure~\ref{fig:merger_quan}). Over time, an increasingly large number of particles that initially belonged to GC 77 are able to gain sufficient energy to unbind themselves from its gravitational field. GC 73, being more massive, does not experience significant mass loss, but it attracts and accretes the majority of the particles lost from GC 77 (see Section~\ref{mass}). By $9.75 \Gyr$, it is no longer possible to identify a bound collection of particles for GC 77: all particles that initially belonged to GC 77 are either unbound ($13 \%$) or are bound to GC 73 ($87 \%$), and the two GCs are said to have merged. Note the strongly elongated nature of the final merged remnant.

\section{Discussion} \label{discussion}

Why is it that we find such a low GC-GC merger rate in our simulations (only 1 out of a total of $3 \times 45$ GC pairs\footnote{Each simulation has $\binom{10}{2} = \frac{10!}{2!8!}=45$ GC pairs. Therefore, in total, the three live-GC simulations have $3 \times 45$ pairs of GCs.} merge in a period of $10 \Gyr$) even though the velocity dispersion of the GC system, $\sigma_{\rm 3D, sys}$, is similar to the intrinsic dispersion of the GCs, $\sigma_{\rm 3D}$? 

To answer this question, we need to ask what it takes for two GCs to merge. First of all, the GCs need to become bound to each other, thus forming a GC-binary. Therefore, we begin our discussion by defining the boundedness condition for a pair of GCs and identifying the various mechanisms by which they can become bound to each other. Next, we discuss tidal shock capture due to impulsive encounters, which is typically considered to be the main mechanism driving the merging of galaxies in groups and clusters, and show that its rate is too low to be relevant for DF2. We end by estimating the average merger rate of the GCs in DF2, and demonstrate that it is dominated by two other mechanisms: dissipative capture, driven by dynamical friction, and three-body capture, driven by interactions of a GC pair with one (or more) of the other GCs. 

\subsection{Foundations} \label{sec:found}

Consider two GCs, with masses $m_1$ and $m_2$, moving in an external potential, $\Phi_{\rm ext}(\bR)$. The total orbital energy of the GCs is given by
\begin{eqnarray}
E_{\rm tot} & = & \frac{1}{2} m_1 V^2_1 +  \frac{1}{2} m_2 V^2_2  + W_{12} + \nonumber \\
& & m_1 \Phi_{\rm ext}(\bR_1) + m_2 \Phi_{\rm ext}(\bR_2)\,.
\label{Etot}
\end{eqnarray}
Here, $\bV_i$ and $\bR_i$ are the velocity and position vectors of GC $i$ with respect to the center of the external potential, and $W_{12}$ is their mutual gravitational potential energy, defined in Equation~\ref{pot_def}. It can also be written in the form 
\begin{equation}
W_{12} = -\frac{G \, m_1 \, m_2}{r_{12}} \, \calS(r_{12})\,,
\label{W12}
\end{equation}
where $r_{12} = |\bR_1-\bR_2|$ is the distance between GCs 1 and 2, and $\calS(x)$ is a function that depends on the density profiles of the two GCs and which asymptotes to unity in the limit of large separation. 

If we define the center-of-mass velocity of the GCs as
\begin{equation}
\bV_{\rm cm} = \frac{m_1 \, \bV_1 + m_2 \, \bV_2}{M}\,,  
\label{cov}
\end{equation}
where $M \equiv m_1 + m_2$, then we can rewrite the total energy as
\begin{equation}
E_{\rm tot} = \frac{1}{2} M V^2_{\rm cm} +   m_1 \Phi_{\rm ext}(\bR_1) + m_2 \Phi_{\rm ext}(\bR_2) + E_{12}\,.
\label{Etot_alt}
\end{equation}
Here,
\begin{eqnarray}
E_{12} = \frac{1}{2} \mu v^2_{\rm 12} - \frac{G \, m_1 \, m_2 }{r_{12}} \, \calS(r_{12}) \,
\label{E12}
\end{eqnarray}
specifies the binding energy (or relative orbital energy) of the GC pair, with $\mu = m_1 \, m_2 / M$ and $v_{12} = |\bV_1 -\bV_2|$ their reduced mass and relative speed, respectively. A pair of GCs is considered to be bound if $E_{12} < 0$. 

We emphasize, though, that a bound pair is not guaranteed to remain bound. In particular, as we show in Appendix~\ref{App:A}, if the pair is comprised of hard spheres and the external potential is time-invariant, then as it evolves from a configuration `a' at time $t$ to a configuration `b' at time $t + \rmd t$, the binding energy changes according to
\begin{align}
 \Delta E_{12}(a \rightarrow b) & = \Wtide\ .
 \label{work-energy}
\end{align}
Here, $\Wtide$ is the work done on the GC pair by the external tidal field, which can be either positive or negative, corresponding to a loosening or tightening of the binary, respectively.

Equation~\ref{work-energy} is basically just an expression for the work-energy principle. In fact, in our simulations, where each GC is live and moves in a live external potential, determined by the galaxy and the other GCs, we may replace $\Wtide$ with the {\it total} work, which includes, in addition to $\Wtide$, the work due to a variety of additional processes, which we split into three categories:
\begin{enumerate}
    \item work done due to mutual gravitational interactions between the GCs that make up the pair.
    \item work done due to gravitational interactions between the GC pair and individual particles belonging to the galaxy.
    \item work done due to gravitational interactions between the GC pair and other GCs. 
\end{enumerate}

Here, we are interested in mechanisms that can cause an initially unbound GC pair ($E_{12}>0$) to become bound ($E_{12}<0$). The primary mechanism of category 1 that can do so is a close, impulsive encounter between two GCs, which causes a transfer of relative orbital energy to internal energy of the GCs \citep[][]{spitzer58}. If this energy transfer is sufficiently large, $E_{12}$ can become negative. In what follows, we refer to this as `tidal shock capture' (tidal capture for short). In the literature, tidal shock capture is considered to be the principal driver of mergers among galaxies (or subhalos) in clusters and groups of galaxies \citep[e.g.,][]{richstone75, white78, roos79, mamon92, makino97}. However, there are other effects to be considered as well. For example, the main mechanism belonging to category 2 that can cause a pair of GCs to become bound is dynamical friction. Since different GCs experience different amounts of dynamical friction, by chance, two GCs can become bound to each other. In what follows, we refer to this as `dissipative capture'. Finally, compressive tidal forces acting on a GC pair from one (or more) of the other eight GCs can also cause it to become bound (a category 3 mechanism), and this is hereafter referred to as `three-body capture'. Note that irrespective of how a GC pair becomes bound, to merge, it must subsequently harden as a binary (via mutual tides, dynamical friction, compressive tides from the galaxy, or other GCs) before disruptive tidal forces from the galaxy or other GCs can rip it apart again.

\begin{figure*}
    \centering
    \includegraphics[width=0.95\textwidth]{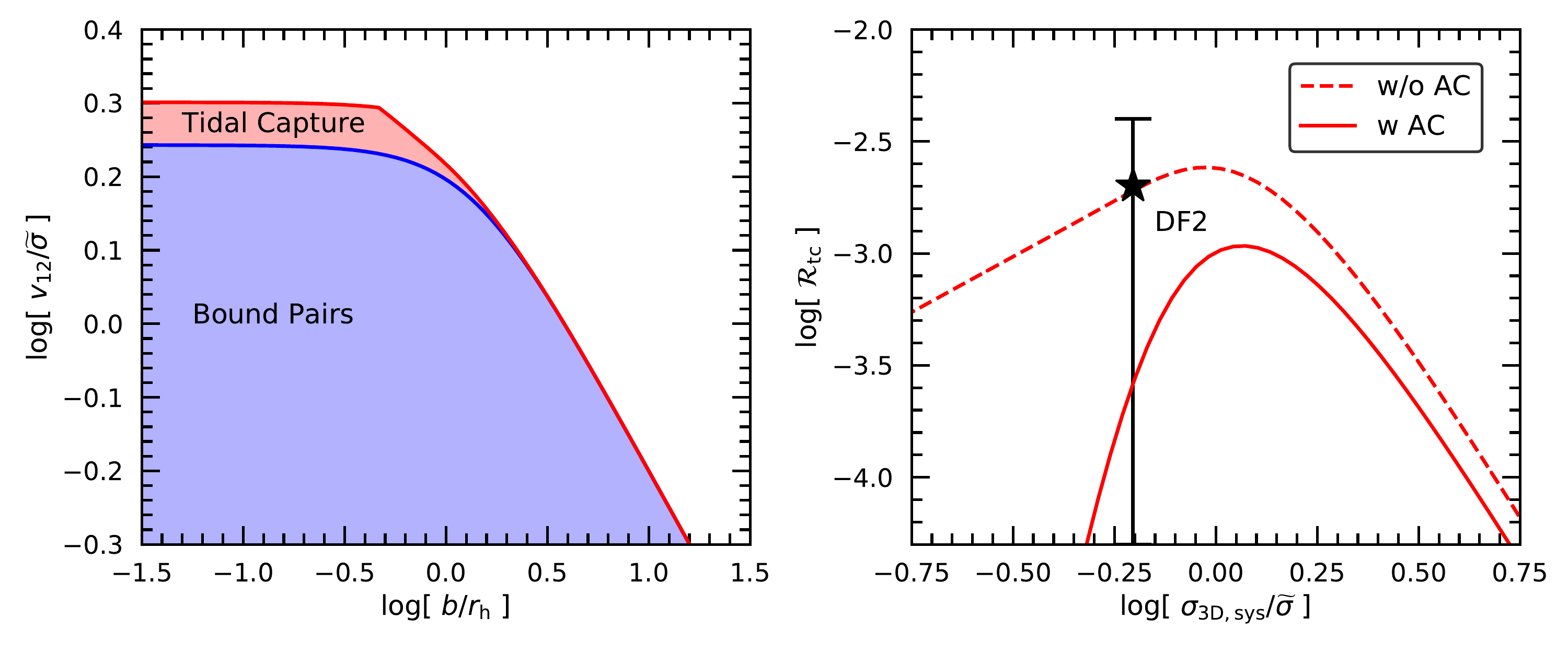}
    \caption{Tidal Shock Capture. For a pair of GCs, modeled as Plummer spheres of identical mass, $M_{\rm GC}=7.2 \times 10^{5} \Msun$, and 3D half-mass radius, $r_\rmh=8.4 \pc$, the blue shaded region in the left-hand panel indicates the range of impact parameters, $b$, and encounter velocities, $v_{12}$, normalized by $r_\rmh$ and  $\tsigma \equiv \sqrt{G M_{\rm GC}/r_\rmh}$, respectively, for which the pair of GCs is bound (i.e., for which $E_{12} \leq 0)$. The red shaded region shows the $b-v_{12}$ parameter space for which an encounter results in tidal shock capture (i.e., for which initial $E_{12} > 0$ and $\Delta E > E_{12}$). Here, $\Delta E$ is the relative orbital energy lost to internal energy during the impulsive encounter and is given by equation~(\ref{dEcomb}). The solid, red curve in the right-hand panel depicts the tidal capture rate, $\calR_{\rm tc}$, in units of $\Gyr^{-1}$, among the GCs in DF2, as a function of $\sigma_{\rm 3D, sys}$, shown in units of $\tsigma$. The dashed, red curve indicates the capture rate obtained if adiabatic shielding of the central regions of the GCs is ignored (see Appendix~\ref{App:B} for details). The black asterisk with error bar indicates the average tidal capture rate inferred from the 50 simulations of Paper~I, as described in the text.}
    \label{fig:tidal_capture}
\end{figure*}

\subsection{Tidal Capture Rate} \label{sec:tsc}

We now proceed to compute the rate at which a pair of GCs, modeled as Plummer spheres of identical mass and scale radius, is expected to undergo tidal shock capture in DF2. This requires that the two GCs have an encounter, characterized by impact parameter, $b = |\bR_1 - \bR_2|$ and encounter velocity, $v_{12} = |\bV_1 - \bV_2|$, that results in a loss of orbital energy, $\Delta E$, that is larger than $E_{12}$ (initially assumed positive), such that post-encounter $E_{12}<0$. As detailed in Appendix~\ref{App:B}, for a given impact parameter, this requires an encounter velocity, $v_{12} < v_{\rm crit}(b)$. The red curve in the left-hand panel of Figure~\ref{fig:tidal_capture} plots this critical velocity, expressed in units of $\tsigma \equiv \sqrt{G M_{\rm GC}/r_\rmh}$, which is proportional to $\sigma_{\rm 3D}$, as a function of the unit-less impact parameter, $\tb = b/r_\rmh$. At large impact parameters ($b \gg r_\rmh$), the critical velocity scales as $v_{\rm crit} \propto b^{-1/2}$. At smaller impact parameters, the detailed density profiles of the GCs cause the critical velocity to asymptote to a finite value as $b \rightarrow 0$ (corresponding to a head-on encounter).

The blue curve in the left-hand panel corresponds to $v_{\rm bound}(b)$, defined as the encounter velocity at a given impact parameter for which $E_{12}=0$.  Using the definition of the binary's binding energy, it is easy to see that $v_{\rm bound}$ is the root for $v_{12}$ of
\begin{equation}
\left( \frac{v_{12}}{\tsigma} \right)^2 = 4 \, \frac{\calS(r_{\rmh} \tb)}{\tb} \ ,
\label{v_bound}
\end{equation}
which for large impact parameters asymptotes to $v_{12} = 2\,\tsigma\,\tb^{-1/2}$. The red shaded region indicates the parameter space of impact parameter and encounter velocity that results in a tidal shock capture. The blue shaded region, on the other hand, represents encounters with $v_{12}<v_{\rm bound}(b)$, which occur between GC pairs that are already bound. Note that tidal shock capture basically requires an impact parameter, $b\lta r_\rmh$ and a very restricted range of encounter velocities. Hence, we expect it to be rare.

In order to quantify this better, we define the rate at which a single GC undergoes tidal shock capture with other GCs as $\Gamma_{\rm tc} = n \, \langle \sigma_{\rm tc}(v_{12}) \, v_{12} \rangle$. Here, $n = 3 N / 4 \pi R_{\rm sys}^3$ is the (approximate) number density of $N$ GCs distributed within a sphere of radius, $R_{\rm sys}$, $\sigma_{\rm tc}(v_{12})$ is the velocity dependent cross-section for tidal shock capture, and the angle brackets indicate an averaging over the encounter velocities, $v_{12}$. If we define $b_{\rm max}$ and $b_{\rm min}$ as the impact parameters for which $v_{12} = v_{\rm crit}(b)$ and $v_{12} = v_{\rm bound}(b)$, respectively, then we have $\sigma_{\rm tc}(v_{12}) = \pi [b^2_{\rm max}(v_{12}) - b^2_{\rm min}(v_{12})]$. We assume that the encounter velocities follow a Maxwell Boltzmann distribution, such that
\begin{equation}
f(v_{12}) \, \rmd v_{12} = \sqrt{\frac{2}{\pi}} \, \frac{v_{12}^2}{\sigma_v^3}  \, \exp\left[ \frac{-v_{12}^2} {2\sigma^2_v}\right]\, \rmd v_{12} \ ,
\label{maxwell_boltzmann}
\end{equation}
where $\sigma_v = \sqrt{2/3} \, \sigma_{\rm 3D, sys}$ is the dispersion in encounter velocities. This yields a tidal capture rate, $\calR_{\rm tc} = N \, \Gamma_{\rm tc}$, given by
\begin{align}
\calR_{\rm tc} & = 0.5984 \Gyr^{-1} \, N^2 \, \left( \frac{t_{\rm cross}}{ \Gyr}\right)^{-1} \nonumber \\
 &  \times \left(\frac{r_\rmh}{R_{\rm sys}}\right)^3 \,  \left(\frac{\sigma_v}{ \tsigma}\right)^{-3} \, \Theta\left(\frac{\sigma_v}{ \tsigma}\right) \ ,
\label{R_tc}
\end{align}
with  $t_{\rm cross} \equiv r_\rmh/\tsigma$, a rough measure of the average crossing time inside a GC, and
\begin{equation}
\Theta(x) = \int_0^{\infty} \rmd y \, y^3 \, \rme^{-\frac{y^2}{2 x^2}} \, \left[\tb^2_{\rm max}(y) - \tb^2_{\rm min}(y) \right]\,.
\label{theta}
\end{equation}

To apply this to DF2's GC system, where $N=10$, we use the masses and scale radii of the GCs listed in Table~1 to obtain an average GC mass of $M_{\rm GC}=7.2 \times 10^5 \Msun$ and an average 3D half-mass radius of $r_\rmh = 8.4 \pc$. This implies that $\tsigma = 19.2 \kms$, and thus $t_{\rm cross} = 0.43 \Myr$. For the size of the entire GC system, we adopt $R_{\rm sys} = R_{\rm half,GC} = 2.86 \kpc$.  The resulting tidal capture rate as a function of $\sigma_{\rm 3D, sys}$, plotted in units of $\tsigma$, is shown by the solid, red curve in the right-hand panel of Figure~\ref{fig:tidal_capture}. For comparison, the dashed, red curve indicates the tidal capture rate obtained if one ignores adiabatic shielding of the central regions of the GCs (see Appendix~\ref{App:B} for details).

In order to estimate the actual tidal capture rate in our simulations, we use all 50 simulations from Paper~I, where the GCs are modeled as hard spheres. For each pair of GCs, in each simulation, we compute the impact parameter, $b=|\bR_1 - \bR_2|$ and encounter velocity, $v_{12}=|\bV_1 - \bV_2|$ during every encounter, defined as a point in time when the dot product $(\bR_1 - \bR_2) \cdot (\bV_1 - \bV_2) = 0$. The tidal capture rate in a particular simulation is then defined as the total number of encounters for which $b$ and $v_{12}$ fall in the red region of the left-hand panel of Figure~\ref{fig:tidal_capture}, divided by the total run time of 10 Gyr. We only find one capture in one out of 50 simulations, which implies an average tidal capture rate of $\calR_{\rm tc} = 0.002 \pm 0.002\Gyr^{-1}$, where the error is determined using the jackknife method. This is indicated by the black asterisk with error bar in the right-hand panel of Figure~\ref{fig:tidal_capture}, where we have adopted $\sigma_{\rm 3D, sys} = 12 \kms$ for DF2 (see Figure~1). Note that the average tidal capture rate inferred from our simulations is in good agreement with the analytically predicted value. 

We emphasize that there are no tidal capture events among the three simulations with live GCs presented in this paper. In particular, the one and only merger in these simulations (between GCs 73 and 77 in Run B) results from capture due to dissipative work done by dynamical friction. That this merger does not result from a tidal capture is clear from the lower panel of Figure~\ref{fig:merger_quan}, where the evolution of $E_{12}$ in Run B (live GCs) starts to deviate from that in Run B' (single-particle GCs) only after the two GCs have become bound. Therefore, the decrease in $E_{12}$ that leads to the capture is identical for the live and single-particle GCs, and it cannot happen via mutual tides as such interactions are irrelevant for single-particles.

\begin{figure*}
    \centering
    \includegraphics[width=0.9\textwidth]{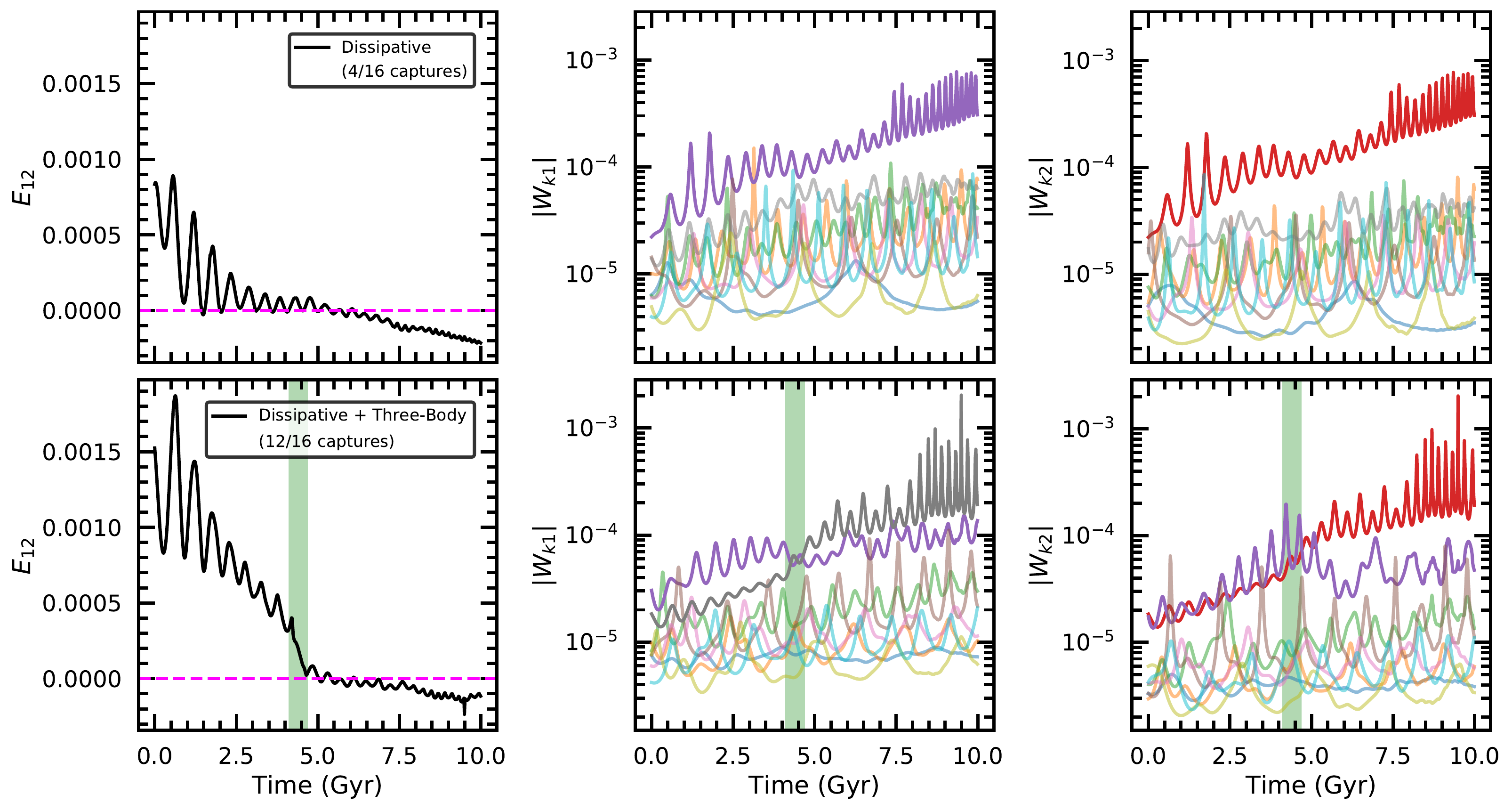}
    \caption{An example each of dissipative (top row) and dissipative $+$ three-body capture (bottom row). Left-hand panels indicate the binding energy, $E_{12}$, of the pair undergoing capture, while the middle and right-hand panels show the magnitudes of the gravitational potential energies of each of these GCs with respect to the other 9 GCs. The capture depicted in the top row results from a steady loss of $E_{12}$ due to dynamical friction, without any significant interactions with other GCs. This is evident from the slow decline of $E_{12}$ and the fact that the magnitude of the gravitational potential energy between the two GCs in question (magenta and red curves in the top middle and right-hand panels, respectively) is always comfortably larger than that of the two GCs with respect to the other eight GCs. In the capture depicted in the bottom row, however, there is a sudden drop in $E_{12}$ from $4.1$ to $4.6\Gyr$ (indicated by the green, vertical band), which coincides with the epoch during which the pair (gray and red curves in bottom middle and right-hand panels, respectively) undergoes strong gravitational interactions with a third GC (the magenta curves).}
    \label{fig:merger_examples}
\end{figure*}

\subsection{Average Merger Rate} \label{sec:merger_rate}

In order to estimate the merger rate among DF2's GCs, we go back to the 50 simulations of Paper~I and identify all GC pairs that undergo a capture and remain bound for a time interval, $\Delta t \ge t_{\rm cross, sys}$, where $t_{\rm cross, sys}=\sqrt{3} R_{\rm sys}/\sigma_{\rm 3D, sys}=0.4 \Gyr$ is the average crossing time of the GC system in DF2. We find 23 such pairs, but remove 7 of them upon `by-eye' inspection. The removed pairs generally have an increasing trend in $E_{12}$ for the greater part of the time interval during which they are bound. Therefore, disruptive tidal forces from the galaxy, other GCs, or both are likely to unbind them before they can merge. The remaining 16 pairs have a decreasing trend in $E_{12}$ for most of the post-capture time interval. As these pairs harden over time, mutual tides (not accounted for in the hard sphere simulations) can be expected to play an increasingly dominant role, transferring relative orbital energy to internal energy of the GCs, eventually leading to a merger, if the GCs were live. Therefore, we estimate that if we had run all 50 Paper~I simulations with live GCs for $10 \Gyr$, we would have found a total of 16 GC-GC mergers corresponding to an average merger rate of $\calR=0.032 \pm 0.007\Gyr^{-1}$, where the error is determined using the jackknife method. We emphasize that the one merger identified in the three live-GC simulations presented here (see Section~\ref{merger}) is correctly identified as a potential merger using this approach (i.e., it is one of the 16 pairs). In fact, from the three live-GC simulations, we obtain an average merger rate of $1/(30 \times 10 \Gyr)=0.033\ \rm Gyr^{-1}$, which is in perfect agreement with the rate inferred from the 50 hard sphere simulations.

Of the 16 captures that are likely to result in a GC-GC merger, 11 are between GCs 73 and 77, 2 between GCs 73 and 92, 2 between GCs 71 and 73, and 1 between GCs 73 and 85. These captures can be broadly classified into two categories - (i) dissipative capture (due to dynamical friction) and (ii) dissipative $+$ three-body capture (due to both dynamical friction and compressive tides from one or more of the other eight GCs). Upon detailed inspection of all 16 cases, we classify 4 as dissipative (which includes the capture between GCs 73 and 77 in Run B) and 12 as dissipative $+$ three-body. Figure~\ref{fig:merger_examples} shows an example each from both categories. Left-hand panels show the binding energy, $E_{12}$, of the pair undergoing capture (referred to as GCs 1 and 2). The middle and the right-hand panels show the magnitudes of the gravitational potential energies of GCs 1 and 2 with respect to the other 9 GCs, respectively.

The capture depicted in the top row is an example of dissipative capture. It takes place between GCs 73 (GC 1) and 77 (GC 2), in a simulation that is different from Run B. In this case, the magnitude of the gravitational potential energy between the two GCs (magenta and red curves in the top middle and right-hand panels, respectively) is always comfortably larger than that of GCs 1 and 2 with respect to the other eight GCs. Therefore, the evolution of the two GCs is unaffected by a third GC, and the decreasing trend in $E_{12}$ (black curve in the top left-hand panel) is entirely due to dynamical friction. At around $5 \Gyr$, this causes $E_{12}$ to become negative, resulting in a capture.

The capture depicted in the bottom row is an example of dissipative $+$ three-body capture. It takes place between GCs 73 (GC 1) and 92 (GC 2). In this case, the motion of the two GCs is affected by a third GC, namely GC 77. This is evident from the bottom left-hand panel, which shows a sudden drop in $E_{12}$ between $4.1$ to $4.6 \Gyr$ (indicated by the green, vertical band). During this time interval, the magnitude of the gravitational potential energy between GCs 73 and 77 (magenta curve in the bottom middle panel) and that between GCs 92 and 77 (magenta curve in the bottom right-hand panel) is similar to the magnitude of the gravitational potential energy between the two GCs (gray and red curves in the bottom middle and right-hand panels, respectively). Therefore, both dynamical friction and compressive tides from GC 77 are responsible for capturing this pair.

\section{Summary} \label{summary}

Dark matter deficient, GC-rich galaxies such as NGC 1052-DF2 present a unique environment for GC evolution. Assuming a baryon-only model for the galaxy, we have studied the evolution of its GCs by modeling them as live $N$-body systems. This study is an improvement over the analysis presented in Paper~I, where the GCs were modeled as hard spheres. It allows us to investigate the occurrence of GC-GC mergers and the impact of the tidal field of DF2 and that of the other GCs on the mass and structural evolution of a GC.

Each GC is initially set up as a spherically symmetric, isotropic Plummer sphere in equilibrium, matching the observational constraints on its mass and projected half-light radius. The projected galactocentric positions and LOS velocities of the GCs are also set in accordance with observations. Their galactocentric positions along the LOS and velocities perpendicular to the LOS are sampled from a DF obtained by assuming the GC system to be in equilibrium with the galaxy. Out of 50 such realizations, the same as the ones simulated in Paper~I, those corresponding to the $16^{\rm th}$, $50^{\rm th}$, and $84^{\rm th}$ percentiles of the probability distribution for the 3D velocity dispersion of the GC system are re-simulated here with live GCs. In order to compare the orbital evolution of the live GCs with that of the corresponding hard spheres, the same three realizations are also re-simulated with single-particle GCs. In each of these simulations, the galaxy is set up as a live, spherically symmetric, isotropic system in equilibrium, matching the observational constraints on its surface brightness profile.

Our findings from these simulations and the accompanying analytical modeling can be summarized as follows:

\begin{itemize}
    \item The GC orbits decay over time due to dynamical friction. However, the amount of orbital decay varies from GC to GC and from realization to realization. Furthermore, even those GCs that experience maximum orbital decay never sink to the galactic center, as reduced dynamical friction in the galactic core (core-stalling) and GC-GC interactions keep them afloat. Thus, the results obtained in Paper~I with single-particle GCs are confirmed here.
    
    \item In the case of live GCs, in addition to providing buoyancy, GC-GC interactions transfer relative orbital energy of a GC pair to internal energy of the GCs, causing their orbits to deviate from that in the corresponding simulations with single-particle GCs. This becomes more important in the later stages of the evolution ($t \gtrsim 5 \Gyr$), after initial orbital decay brings the GCs closer together. During the early stages ($t \lesssim 5 \Gyr$), when they are still far apart, the orbital evolution of live and single-particle GCs are in very good agreement. 
    
    \item Mass loss and structural changes induced by galactic tides are insignificant. The tidal field of one GC on another is also not strong enough to cause significant mass and structural evolution, except when a pair of GCs is about to undergo a merger.
    
    \item Even though the internal velocity dispersion of the GCs is similar to the velocity dispersion of the GC system, GC-GC mergers are rare. In only one of the three live-GC simulations, a single, complete merger occurs within $10 \Gyr$. The merger takes place near the galactic core, where the two GCs congregate after having experienced orbital decay. The merged remnant has a strongly elongated structure and continues to orbit near the core radius of the galaxy.
\end{itemize}

Before two GCs can merge, they must first become bound to one another and subsequently harden as a binary. In DF2, two GCs can become bound via one or more of the following mechanisms: mutual tides (tidal shock capture), dynamical friction (dissipative capture), and compressive tides from one (or more) of the other eight GCs (three-body capture). Using our simulations and analytic modeling, we have shown that tidal shock capture in DF2 is extremely rare (at least within $10 \Gyr$), with an expected rate of $0.002 \pm 0.002 \Gyr^{-1}$. This is because, given their number density, the GCs are too dense for tidal shock capture to operate efficiently. The tidal capture rate is an order of magnitude lower than the inferred merger rate of $0.032 \pm 0.007 \Gyr^{-1}$, which is, therefore, dominated by dissipative and three-body capture. Using the 50 single-particle GC simulations of Paper~I, we estimate that about one-quarter of the expected mergers are driven purely by dissipative capture, with the remaining 75 percent involving both dynamical friction and three-body capture. It is also worth emphasizing that we expect virtually zero mergers during the first $5 \Gyr$ of evolution; most mergers occur at later times once several GCs have congregated close to the core radius of DF2. Note that in the absence of core-stalling (i.e., if DF2 were to have a cuspy dark matter halo), all these GCs would likely end up merging at the center of the galaxy, thus forming a nuclear star cluster.

During a merger, mutual gravitational interactions may result in observable tidal features around one or both of the involved GCs, although such features are not seen for the one merger that occurs in our live-GC simulations. As we have established that mergers can only take place in the later stages of the evolution, and since mass loss due to galactic tides is negligible for the baryon-only mass model that we have assumed for DF2, it is unlikely for the GCs to have tidal features (such as extended tidal tails) in their currently observed state. However, if DF2 were to have a sufficiently dense dark matter halo, depending on their orbit, the GCs may experience significant tidal mass loss. Hence, the presence or absence of tidal features around DF2's GCs can constrain possible mass models. 

As to the origin of the unusually high GC masses, we conclude that it is most likely not an outcome of past GC-GC mergers. If anything, as GC orbits decay over time, the GC system is expected to have been more extended in the past, resulting in an even lower merger rate than inferred here, unless the total number of initial GCs was much higher than that at present. For instance, if DF2 initially had $\sim 100$ GCs, roughly distributed within the same volume as the current population, its past tidal capture rate would have been higher by a factor of $100$ (see Equation~\ref{R_tc}). As discussed in \citet{leigh20}, this could have resulted in significant evolution of the initial GC luminosity function. However, given that the total number of GCs in DF2 at present is already anomalously high for its stellar mass, postulating an even larger number of GCs in the past seems a bit far-fetched. Hence, the abundance and luminosity function of the GCs in DF2 continues to be an enigma for our current understanding of galaxy formation.

\acknowledgments

The authors are grateful to Uddipan Banik, Nir Mandelker, Victor Robles, and Zili Shen for valuable discussions and to the anonymous referee for insightful feedback. DDC thanks the Yale Center for Research Computing for guidance and use of the research computing infrastructure, specifically the Grace cluster. FvdB is supported by the National Aeronautics and Space Administration through Grant Nos. 17-ATP17-0028 and 19-ATP19-0059 issued as part of the Astrophysics Theory Program and received additional support from the Klaus Tschira foundation. 


\appendix

\section{Tidal Evolution of the Binding Energy of a Pair of Globular Clusters} \label{App:A}

In Section~\ref{discussion}, we defined the binding energy of a pair of GCs as 
\begin{equation}
E_{12} = \frac{1}{2}\, \mu \, v^2_{\rm 12} - \frac{G \, m_1 \, m_2}{r_{12}} \, \calS(r_{12}) \,,
\end{equation}
with $\mu = (m_1 \, m_2) / M$ the reduced mass, $\bv_{12} = \bV_1 -\bV_2$ their relative velocity, $\br_{12} = \bR_1 -\bR _2$ their relative separation, and $\calS(x)$ a function that depends on the density profiles of the two GCs. Also, $\bR_{i}$ and $\bV_{i}$ are the position and velocity vectors of GC $i$ with respect to the center of the external potential. Under the assumption that the GCs are hard spheres and the external potential is time-invariant, we now derive an expression for how $E_{12}$ evolves as the GCs orbit in this potential. 

Let $\br'_1$ and $\br'_2$ be the position vectors of GCs 1 and 2 from their common center-of-mass, whose position vector with respect to the center of the external potential is $\bR_{\rm cm }$. Then
\begin{equation}
    \br'_1 = \bR_1 - \bR_{\rm cm}\,,\;\;\;\;\;\;\;\;\;\;\;\;
    \br'_2  = \bR_2 - \bR_{\rm cm}\,.
\end{equation}
Applying Newton's second law of motion in the center-of-mass frame of the two GCs, we have that
\begin{equation}
m_1 \, \ddot{\br}'_1 = \bF_{12} + \bF_{\rm ext}(\bR_1) - m_1 \, \bA_{\rm cm}\,, \;\;\;\;\;\;\;\;\;\;\;\;
m_2 \, \ddot{\br}'_2 = \bF_{21} + \bF_{\rm ext}(\bR_2) - m_2 \, \bA_{\rm cm}\,,
\end{equation}
where $\bF_{ij}$ is the gravitational force due to GC $j$ on GC $i$, $\bF_{\rm ext}(\bR)$ is the force at $\bR$ due to the external potential, and $\bA_{\rm cm} = \rmd^2 \bR_{\rm cm}/\rmd t^2$ is the acceleration of the center-of-mass. The force, $-m_{i}\bA_{\rm cm}$ acting on GC $i$ is a pseudo-force, which comes into play because the center-of-mass frame of the two GCs is non-internal.

Let $\rmd \br'_i$ correspond to the displacement vector of GC $i$ in time interval $\rmd t$ in this frame. Since $\bF_{12} = -\bF_{21}$, we have that
\begin{equation}\label{acca}
 m_1 \, \ddot{\br}'_1 \cdot \rmd\br'_1 + m_2 \, \ddot{\br}'_2 \cdot \rmd\br'_2 =
 \bF_{12} \cdot \rmd \br_{12} + \bF_{\rm ext}(\bR_1) \cdot \rmd\br'_1 + \bF_{\rm ext}(\bR_2) \cdot \rmd\br'_2\,,
\end{equation}
where $\rmd \br_{12} \equiv \rmd \br'_1 - \rmd \br'_2$ is the relative displacement vector of the two GCs, and we have used the fact that
\begin{equation}
\bA_{\rm cm} \cdot \left(m_1 \rmd\br'_1 + m_2 \rmd\br'_2\right) = 0\,,    
\end{equation}
which follows from the definition of the center-of-mass velocity of the two GCs.

Defining $\ddot{\br}'_i = \rmd\bv'_i/\rmd t$, $\rmd\br'_i = \bv'_i \, \rmd t$, and integrating equation~(\ref{acca}) from configuration `a' at time $t$ to configuration `b' at time $t+\rmd t$, one easily obtains that
\begin{equation}\label{accb}
 \frac{1}{2} \, \mu \, v^2_{12,\rmb} - \frac{1}{2} \, \mu \, v^2_{12,\rma} - \int_{a}^{b} 
 \bF_{12} \cdot \rmd \br_{12} = \int_{a}^{b} \bF_{\rm ext}(\bR_1) \cdot \rmd\br'_1
+  \int_{a}^{b} \bF_{\rm ext}(\bR_2) \cdot \rmd\br'_2\,.
\end{equation}
Finally, since gravity is a conservative force, $-\int_a^b \bF_{12} \cdot \rmd \br_{12}$ is simply the difference in mutual gravitational potential energy between configurations $a$ and $b$, such that equation~\ref{accb} reduces to
\begin{equation}
\Delta E_{12}(a \rightarrow b) = \Wtide \equiv \int_{a}^{b} \bF_{\rm ext}(\bR_1) \cdot \rmd\br'_1 +  \int_{a}^{b} \bF_{\rm ext}(\bR_2) \cdot \rmd\br'_2\,,
\end{equation}
where $\Wtide$ is the work done by the external potential on the GC pair.

\section{Tidal Shock Capture} \label{App:B}

Consider an impulsive encounter between two GCs with impact parameter, $b$ and encounter velocity, $v_{12}$. In what follows, for simplicity, we assume that all GCs are Plummer spheres with identical mass, $m$, and half-mass radius, $r_\rmh$. In the distant tide approximation ($b \gg r_\rmh$), following \citet{spitzer58} and \cite{gnedin99a}, the loss in relative orbital energy as a result of the encounter is given by
\begin{equation}
  \Delta E(b,v_{12}) = \Delta E_1 + \Delta E_2 = \frac{8}{3} \frac{G^2 \, m^3}{v_{12}^2} \, \frac{r^2_\rmh}{b^4} \, \alpha^2(b,v_{12}) \, \chi(b)\,.
  \label{dEimpulse}
\end{equation}
Here, $\alpha$ is a structural parameter given by
\begin{equation}\label{alpha}
 \alpha^2 \equiv  \frac{\langle r^2 \rangle_{\rm AC}}{r^2_\rmh} = \frac{4 \pi}{m r^2_\rmh} \int_0^{r_\rmv} \rho(r) \, f_{\rm AC}(r) \, r^4 \, \rmd r\,,
\end{equation}
with $\rho(r)$ the density profile of the GCs and $f_{\rm AC}(r)$ a function, detailed below, that accounts for the fact that the central regions of the GCs may be adiabatically shielded \citep[e.g.,][]{weinberg94, gnedin99b}. Note that the integral is truncated at the virial radius, $r_\rmv$\footnote{This truncation is required, since the integral on the RHS of equation~\ref{alpha} diverges otherwise.}, defined such that the self gravitational potential energy of a GC is given by $W=-G m^2/2 r_\rmv$ \citep[see e.g.,][]{makino97}. For the Plummer spheres considered here, $r_\rmv = 1.3\,r_\rmh$. Finally, the function $\chi(b)$ accounts for the fact that GCs are not point-masses and is given by
\begin{equation}
 \chi(b) = \frac{1}{2} [ (3J_0 - J_1 - I_0)^2 + (2I_0 - I_1 - 3J_0 + J_1)^2 + I_0^2 ]\,,
\end{equation}
where
\begin{equation}
  I_0(b) = \int_1^{\infty} \frac{m(b\zeta)}{m} \, \frac{\rmd \zeta}{\zeta^2 \, (\zeta^2 - 1)^{1/2}}\,, \;\;\;\;\;\;\;\;\;\;\;\;\;\;\;\;
  J_0(b) = \int_1^{\infty} \frac{m(b\zeta)}{m} \, \frac{\rmd \zeta}{\zeta^4 \, (\zeta^2 - 1)^{1/2}}\,,
\end{equation}
with $m(r)$ the GC mass enclosed within cluster-centric radius $r$, and $I_1(b) = b \, \rmd I_0/\rmd b$ and $J_1(b) = b \, \rmd J_0/\rmd b$ \citep[][]{gnedin99a}. 

Equation~\ref{dEimpulse} is only valid for relatively distant encounters with $b \gg r_\rmh$. In order to obtain an expression for $\Delta E(b,v_{12})$ that is valid for all $b$, we follow \cite{vandenbosch18} and set
\begin{equation}\label{dEcomb}
  \Delta E(b,v_{12}) = \frac{8 G^2 m^3}{3 v_{12}^2} \, r^2_\rmh \, \left\{
  \begin{array}{ll}
     \alpha^2(b,v_{12}) \, \frac{\chi(b)}{b^4} & \mbox{if $b > b_0$} \\
     \alpha^2(b_0,v_{12}) \frac{\chi(b_0)}{b_0^4} & \mbox{if $b\leq b_0$} 
  \end{array}
  \right..
\end{equation}
Here, $b_0$ is defined as the impact parameter for which $\Delta E$ obtained using equation~(\ref{dEimpulse}) is equal to that of a head-on ($b=0$) collision, which is given by
\begin{equation}\label{headon}
\Delta E_0 = \frac{8 G^2 m^2}{v_{12}^2} \, \pi \, \int_0^{r_\rmv} I_0^2(r) \, \Sigma_\rms(r) \, \frac{\rmd r}{r}\,,
\end{equation}
with $\Sigma_\rms(r)$ the projected surface density profile of the GCs. Hence, at small $b$, we assume that $\Delta E(b,v_{12})$ is equal to that of a head-on encounter. As shown in Banik \& van den Bosch (2020, in prep.), this accurately captures the dependence of $\Delta E$ over the full range of impact parameters.

Since the encounter velocities among the GCs are comparable to the internal velocities of the GCs, the encounters are only marginally impulsive. Hence, it is important to correct for adiabatic
shielding. We follow \cite{gnedin99a} and adopt\footnote{Ongoing studies suggest that this treatment may require a revision for the case of extensive tides (O. Gnedin, private communication).}
\begin{equation}
f_{\rm AC}(r) = \left[1 + \omega^2(r) \tau^2\right]^{-\gamma}\,.
\end{equation}
Here, $\tau = b/v_{12}$ is the duration of the impulsive shock, and $\omega(r) = v_\rmc(r)/r$ is the angular velocity for a circular orbit at radius $r$, with $v_\rmc(r) = \sqrt{G m(r)/r}$ the circular speed. The value of $\gamma$ increases from 1.5 for relatively slow encounters with $\tau > 4 t_{\rm dyn}$ to 2.5 for fast encounters with $\tau < t_{\rm dyn}$. Here, $t_{\rm dyn}$ is the half-mass dynamical time of the GC. 

Tidal shock capture occurs if prior to the encounter, $E_{12} > 0$, and the encounter results in a $\Delta E > E_{12}$. For a given impact parameter, $b$, this requires an encounter speed, $v_{12} < v_{\rm crit}(b)$, which is given by the root for $v_{12}$ of $\Delta E(b,v_{12}) = E_{12}$. We use a simple root-finder to numerically compute $v_{\rm crit}(b)$. The result for encounters among two identical Plummer spheres is shown as the red, solid line in the left-hand panel of Figure~\ref{fig:tidal_capture}. Note that this critical encounter velocity is expressed in units of $\tsigma \equiv \sqrt{G m/r_\rmh}$, which is proportional to the internal velocity dispersion of the GCs, while the impact parameter, $b$, is expressed in units of the GC's half-mass radius, $r_\rmh$.

\bibliography{ms}

\begin{thebibliography}{}
\expandafter\ifx\csname natexlab\endcsname\relax\def\natexlab#1{#1}\fi
\providecommand{\url}[1]{\href{#1}{#1}}
\providecommand{\dodoi}[1]{doi:~\href{http://doi.org/#1}{\nolinkurl{#1}}}
\providecommand{\doeprint}[1]{\href{http://ascl.net/#1}{\nolinkurl{http://ascl.net/#1}}}
\providecommand{\doarXiv}[1]{\href{https://arxiv.org/abs/#1}{\nolinkurl{https://arxiv.org/abs/#1}}}

\bibitem[{{Arca-Sedda} \& {Capuzzo-Dolcetta}(2014)}]{arca-sedda14}
{Arca-Sedda}, M., \& {Capuzzo-Dolcetta}, R. 2014, \mnras, 444, 3738,
  \dodoi{10.1093/mnras/stu1683}

\bibitem[{{Barnes} \& {Hut}(1986)}]{barnes86}
{Barnes}, J., \& {Hut}, P. 1986, \nat, 324, 446, \dodoi{10.1038/324446a0}

\bibitem[{{Bekki}(2010)}]{bekki10}
{Bekki}, K. 2010, \mnras, 401, 2753, \dodoi{10.1111/j.1365-2966.2009.15874.x}

\bibitem[{{Bekki} {et~al.}(2004){Bekki}, {Couch}, {Drinkwater}, \&
  {Shioya}}]{bekki04}
{Bekki}, K., {Couch}, W.~J., {Drinkwater}, M.~J., \& {Shioya}, Y. 2004, \apjl,
  610, L13, \dodoi{10.1086/423130}

\bibitem[{{Binney} \& {Tremaine}(2008)}]{binney08}
{Binney}, J., \& {Tremaine}, S. 2008, {Galactic Dynamics: Second Edition}
  (Princeton University Press)

\bibitem[{{Blakeslee} \& {Cantiello}(2018)}]{blakeslee18}
{Blakeslee}, J.~P., \& {Cantiello}, M. 2018, Research Notes of the American
  Astronomical Society, 2, 146, \dodoi{10.3847/2515-5172/aad90e}

\bibitem[{{Capuzzo-Dolcetta} \&
  {Miocchi}(2008{\natexlab{a}})}]{capuzzo-dolcetta08a}
{Capuzzo-Dolcetta}, R., \& {Miocchi}, P. 2008{\natexlab{a}}, \apj, 681, 1136,
  \dodoi{10.1086/588017}

\bibitem[{{Capuzzo-Dolcetta} \&
  {Miocchi}(2008{\natexlab{b}})}]{capuzzo-dolcetta08b}
---. 2008{\natexlab{b}}, \mnras, 388, L69,
  \dodoi{10.1111/j.1745-3933.2008.00501.x}

\bibitem[{{Danieli} {et~al.}(2019){Danieli}, {van Dokkum}, {Conroy}, {Abraham},
  \& {Romanowsky}}]{danieli19}
{Danieli}, S., {van Dokkum}, P., {Conroy}, C., {Abraham}, R., \& {Romanowsky},
  A.~J. 2019, \apjl, 874, L12, \dodoi{10.3847/2041-8213/ab0e8c}

\bibitem[{{Dutta Chowdhury} {et~al.}(2019){Dutta Chowdhury}, {van den Bosch},
  \& {van Dokkum}}]{duttachowdhury19}
{Dutta Chowdhury}, D., {van den Bosch}, F.~C., \& {van Dokkum}, P. 2019, \apj,
  877, 133, \dodoi{10.3847/1538-4357/ab1be4}

\bibitem[{{Emsellem} {et~al.}(2019){Emsellem}, {van der Burg}, {Fensch},
  {Je{\v{r}}{\'a}bkov{\'a}}, {Zanella}, {Agnello}, {Hilker}, {M{\"u}ller},
  {Rejkuba}, {Duc}, {Durrell}, {Habas}, {Lelli}, {Lim}, {Marleau}, {Peng}, \&
  {S{\'a}nchez-Janssen}}]{emsellem19}
{Emsellem}, E., {van der Burg}, R. F.~J., {Fensch}, J., {et~al.} 2019, \aap,
  625, A76, \dodoi{10.1051/0004-6361/201834909}

\bibitem[{{Fellhauer} \& {Kroupa}(2002)}]{fellhauer02}
{Fellhauer}, M., \& {Kroupa}, P. 2002, \mnras, 330, 642,
  \dodoi{10.1046/j.1365-8711.2002.05087.x}

\bibitem[{{Gnedin} {et~al.}(1999){Gnedin}, {Hernquist}, \&
  {Ostriker}}]{gnedin99a}
{Gnedin}, O.~Y., {Hernquist}, L., \& {Ostriker}, J.~P. 1999, \apj, 514, 109,
  \dodoi{10.1086/306910}

\bibitem[{{Gnedin} \& {Ostriker}(1999)}]{gnedin99b}
{Gnedin}, O.~Y., \& {Ostriker}, J.~P. 1999, \apj, 513, 626,
  \dodoi{10.1086/306864}

\bibitem[{{Gnedin} {et~al.}(2014){Gnedin}, {Ostriker}, \&
  {Tremaine}}]{gnedin14}
{Gnedin}, O.~Y., {Ostriker}, J.~P., \& {Tremaine}, S. 2014, \apj, 785, 71,
  \dodoi{10.1088/0004-637X/785/1/71}

\bibitem[{{Hartmann} {et~al.}(2011){Hartmann}, {Debattista}, {Seth},
  {Cappellari}, \& {Quinn}}]{hartmann11}
{Hartmann}, M., {Debattista}, V.~P., {Seth}, A., {Cappellari}, M., \& {Quinn},
  T.~R. 2011, \mnras, 418, 2697, \dodoi{10.1111/j.1365-2966.2011.19659.x}

\bibitem[{{Hayashi} \& {Inoue}(2018)}]{hayashi18}
{Hayashi}, K., \& {Inoue}, S. 2018, \mnras, 481, L59,
  \dodoi{10.1093/mnrasl/sly162}

\bibitem[{{Hernandez} \& {Gilmore}(1998)}]{hernandez98}
{Hernandez}, X., \& {Gilmore}, G. 1998, \mnras, 297, 517,
  \dodoi{10.1046/j.1365-8711.1998.01511.x}

\bibitem[{{Inoue}(2009)}]{inoue09}
{Inoue}, S. 2009, \mnras, 397, 709, \dodoi{10.1111/j.1365-2966.2009.15066.x}

\bibitem[{{Inoue}(2011)}]{inoue11}
---. 2011, \mnras, 416, 1181, \dodoi{10.1111/j.1365-2966.2011.19122.x}

\bibitem[{{Kaur} \& {Sridhar}(2018)}]{kaur18}
{Kaur}, K., \& {Sridhar}, S. 2018, \apj, 868, 134,
  \dodoi{10.3847/1538-4357/aaeacf}

\bibitem[{{Khoperskov} {et~al.}(2018){Khoperskov}, {Mastrobuono-Battisti}, {Di
  Matteo}, \& {Haywood}}]{khoperskov18}
{Khoperskov}, S., {Mastrobuono-Battisti}, A., {Di Matteo}, P., \& {Haywood}, M.
  2018, \aap, 620, A154, \dodoi{10.1051/0004-6361/201833534}

\bibitem[{{Kroupa}(1998)}]{kroupa98}
{Kroupa}, P. 1998, \mnras, 300, 200, \dodoi{10.1046/j.1365-8711.1998.01892.x}

\bibitem[{{Laporte} {et~al.}(2019){Laporte}, {Agnello}, \&
  {Navarro}}]{laporte19}
{Laporte}, C.~F.~P., {Agnello}, A., \& {Navarro}, J.~F. 2019, \mnras, 484, 245,
  \dodoi{10.1093/mnras/sty2891}

\bibitem[{{Leigh} \& {Fragione}(2020)}]{leigh20}
{Leigh}, N. W.~C., \& {Fragione}, G. 2020, \apj, 892, 32,
  \dodoi{10.3847/1538-4357/ab7a8f}

\bibitem[{{Lewis} {et~al.}(2020){Lewis}, {Brewer}, \& {Wan}}]{lewis20}
{Lewis}, G.~F., {Brewer}, B.~J., \& {Wan}, Z. 2020, \mnras, 491, L1,
  \dodoi{10.1093/mnrasl/slz157}

\bibitem[{{Makino} \& {Hut}(1997)}]{makino97}
{Makino}, J., \& {Hut}, P. 1997, \apj, 481, 83, \dodoi{10.1086/304013}

\bibitem[{{Mamon}(1992)}]{mamon92}
{Mamon}, G.~A. 1992, \apjl, 401, L3, \dodoi{10.1086/186656}

\bibitem[{Martin {et~al.}(2018)Martin, Collins, Longeard, \&
  Tollerud}]{martin18}
Martin, N.~F., Collins, M. L.~M., Longeard, N., \& Tollerud, E. 2018, \apjl,
  859, L5

\bibitem[{{Mastrobuono-Battisti} {et~al.}(2019){Mastrobuono-Battisti},
  {Khoperskov}, {Di Matteo}, \& {Haywood}}]{mastrobuono-battisti19}
{Mastrobuono-Battisti}, A., {Khoperskov}, S., {Di Matteo}, P., \& {Haywood}, M.
  2019, \aap, 622, A86, \dodoi{10.1051/0004-6361/201834087}

\bibitem[{{Nusser}(2018)}]{nusser18}
{Nusser}, A. 2018, \apjl, 863, L17, \dodoi{10.3847/2041-8213/aad6ee}

\bibitem[{{Nusser}(2019)}]{nusser19a}
---. 2019, \mnras, 484, 510, \dodoi{10.1093/mnras/sty3532}

\bibitem[{{Nusser}(2020)}]{nusser20}
---. 2020, \apj, 893, 66, \dodoi{10.3847/1538-4357/ab792c}

\bibitem[{{Ogiya}(2018)}]{ogiya18}
{Ogiya}, G. 2018, \mnras, 480, L106, \dodoi{10.1093/mnrasl/sly138}

\bibitem[{{Oh} \& {Lin}(2000)}]{oh00}
{Oh}, K.~S., \& {Lin}, D.~N.~C. 2000, \apj, 543, 620, \dodoi{10.1086/317118}

\bibitem[{{Petts} {et~al.}(2015){Petts}, {Gualandris}, \& {Read}}]{petts15}
{Petts}, J.~A., {Gualandris}, A., \& {Read}, J.~I. 2015, \mnras, 454, 3778,
  \dodoi{10.1093/mnras/stv2235}

\bibitem[{{Petts} {et~al.}(2016){Petts}, {Read}, \& {Gualandris}}]{petts16}
{Petts}, J.~A., {Read}, J.~I., \& {Gualandris}, A. 2016, \mnras, 463, 858,
  \dodoi{10.1093/mnras/stw2011}

\bibitem[{{Plummer}(1911)}]{plummer11}
{Plummer}, H.~C. 1911, \mnras, 71, 460, \dodoi{10.1093/mnras/71.5.460}

\bibitem[{{Read} {et~al.}(2006){Read}, {Goerdt}, {Moore}, {Pontzen}, {Stadel},
  \& {Lake}}]{read06}
{Read}, J.~I., {Goerdt}, T., {Moore}, B., {et~al.} 2006, \mnras, 373, 1451,
  \dodoi{10.1111/j.1365-2966.2006.11022.x}

\bibitem[{{Richstone}(1975)}]{richstone75}
{Richstone}, D.~O. 1975, \apj, 200, 535, \dodoi{10.1086/153820}

\bibitem[{{Roos} \& {Norman}(1979)}]{roos79}
{Roos}, N., \& {Norman}, C.~A. 1979, \aap, 76, 75

\bibitem[{{S\'ersic}(1968)}]{sersic68}
{S\'ersic}, J.~L. 1968, {Atlas de Galaxias Australes}

\bibitem[{{Shin} {et~al.}(2020){Shin}, {Jung}, {Kwon}, {Kim}, {Lee}, {Jo}, \&
  {Kiat Oh}}]{shin20}
{Shin}, E.-j., {Jung}, M., {Kwon}, G., {et~al.} 2020, arXiv e-prints,
  arXiv:2007.09889.
\newblock \doarXiv{2007.09889}

\bibitem[{{Silk}(2019)}]{silk19}
{Silk}, J. 2019, \mnras, L105, \dodoi{10.1093/mnrasl/slz090}

\bibitem[{{Spitzer}(1958)}]{spitzer58}
{Spitzer}, Lyman, J. 1958, \apj, 127, 17, \dodoi{10.1086/146435}

\bibitem[{{Spitzer}(1969)}]{spitzer69}
---. 1969, \apjl, 158, L139, \dodoi{10.1086/180451}

\bibitem[{{Springel}(2005)}]{springel05}
{Springel}, V. 2005, \mnras, 364, 1105,
  \dodoi{10.1111/j.1365-2966.2005.09655.x}

\bibitem[{{Tremaine} {et~al.}(1975){Tremaine}, {Ostriker}, \&
  {Spitzer}}]{tremaine75}
{Tremaine}, S.~D., {Ostriker}, J.~P., \& {Spitzer}, L., J. 1975, \apj, 196,
  407, \dodoi{10.1086/153422}

\bibitem[{{Trujillo} {et~al.}(2019){Trujillo}, {Beasley}, {Borlaff},
  {Carrasco}, {Di Cintio}, {Filho}, {Monelli}, {Montes}, {Rom{\'a}n},
  {Ruiz-Lara}, {S{\'a}nchez Almeida}, {Valls-Gabaud}, \&
  {Vazdekis}}]{trujillo19}
{Trujillo}, I., {Beasley}, M.~A., {Borlaff}, A., {et~al.} 2019, \mnras, 486,
  1192, \dodoi{10.1093/mnras/stz771}

\bibitem[{{van den Bosch} {et~al.}(2018){van den Bosch}, {Ogiya}, {Hahn}, \&
  {Burkert}}]{vandenbosch18}
{van den Bosch}, F.~C., {Ogiya}, G., {Hahn}, O., \& {Burkert}, A. 2018, \mnras,
  474, 3043, \dodoi{10.1093/mnras/stx2956}

\bibitem[{{van Dokkum} {et~al.}(2019){van Dokkum}, {Danieli}, {Abraham},
  {Conroy}, \& {Romanowsky}}]{vandokkum19}
{van Dokkum}, P., {Danieli}, S., {Abraham}, R., {Conroy}, C., \& {Romanowsky},
  A.~J. 2019, \apjl, 874, L5, \dodoi{10.3847/2041-8213/ab0d92}

\bibitem[{{van Dokkum} {et~al.}(2018{\natexlab{a}}){van Dokkum}, {Danieli},
  {Cohen}, {Romanowsky}, \& {Conroy}}]{vandokkum18d}
{van Dokkum}, P., {Danieli}, S., {Cohen}, Y., {Romanowsky}, A.~J., \& {Conroy},
  C. 2018{\natexlab{a}}, \apjl, 864, L18, \dodoi{10.3847/2041-8213/aada4d}

\bibitem[{{van Dokkum} {et~al.}(2018{\natexlab{b}}){van Dokkum}, {Danieli},
  {Cohen}, {Merritt}, {Romanowsky}, {Abraham}, {Brodie}, {Conroy}, {Lokhorst},
  {Mowla}, {O'Sullivan}, \& {Zhang}}]{vandokkum18a}
{van Dokkum}, P., {Danieli}, S., {Cohen}, Y., {et~al.} 2018{\natexlab{b}},
  \nat, 555, 629, \dodoi{10.1038/nature25767}

\bibitem[{{van Dokkum} {et~al.}(2018{\natexlab{c}}){van Dokkum}, {Cohen},
  {Danieli}, {Kruijssen}, {Romanowsky}, {Merritt}, {Abraham}, {Brodie},
  {Conroy}, {Lokhorst}, {Mowla}, \& {Zhang}}]{vandokkum18b}
{van Dokkum}, P., {Cohen}, Y., {Danieli}, S., {et~al.} 2018{\natexlab{c}},
  \apjl, 856, L30, \dodoi{10.3847/2041-8213/aab60b}

\bibitem[{{van Dokkum} {et~al.}(2018{\natexlab{d}}){van Dokkum}, {Cohen},
  {Danieli}, {Romanowsky}, {Abraham}, {Brodie}, {Conroy}, {Kruijssen},
  {Lokhorst}, {Merritt}, {Mowla}, \& {Zhang}}]{vandokkum18c}
---. 2018{\natexlab{d}}, Research Notes of the American Astronomical Society,
  2, 54, \dodoi{10.3847/2515-5172/aacc6f}

\bibitem[{{Wasserman} {et~al.}(2018){Wasserman}, {Romanowsky}, {Brodie}, {van
  Dokkum}, {Conroy}, {Abraham}, {Cohen}, \& {Danieli}}]{washerman18}
{Wasserman}, A., {Romanowsky}, A.~J., {Brodie}, J., {et~al.} 2018, \apjl, 863,
  L15, \dodoi{10.3847/2041-8213/aad779}

\bibitem[{{Weinberg}(1994)}]{weinberg94}
{Weinberg}, M.~D. 1994, \aj, 108, 1403, \dodoi{10.1086/117162}

\bibitem[{{White}(1978)}]{white78}
{White}, S.~D.~M. 1978, \mnras, 184, 185, \dodoi{10.1093/mnras/184.2.185}

\end{thebibliography}
\bibstyle{aasjournal}

\end{document}